\documentclass[aps,twocolumn, superscriptaddress, footinbib, prr]{revtex4-2}

\usepackage[english]{babel}
\usepackage{lmodern}

\usepackage[dvipsnames]{xcolor}

\usepackage{amsmath, amssymb}
\usepackage{graphicx}
\usepackage[colorlinks=true, allcolors=Maroon]{hyperref}
\usepackage[utf8]{inputenc}
\usepackage{dsfont}
\usepackage[T1]{fontenc}
\usepackage{braket}
\usepackage{mathtools}

\usepackage{algorithm}
\usepackage{algpseudocode}

\usepackage{caption}
\usepackage{tikz}
\usepackage{pgfplots}
\usepackage{pgfplotstable}

\usetikzlibrary{decorations.pathmorphing}
\usetikzlibrary{decorations.pathreplacing,calligraphy}
\usetikzlibrary{decorations.markings}
\usetikzlibrary{calc,fit,patterns}

\pgfplotsset{compat=1.18}
\pgfplotsset{every axis/.append style={
    label style = {font = \large},
    scale = 1,
    grid = major,
    grid style = dashed,
    mark size = 2pt,
}}

\pgfplotsset{every axis plot/.append style = {
    mark=*, 
    only marks,
}}

\tikzstyle{unit}=[circle,draw=red!50,fill=red!20,thick]
\tikzstyle{tensor}=[rectangle,draw=blue!50,fill=blue!20,thick, minimum width=6mm, minimum height=6mm]
\tikzstyle{newTensor}=[rectangle,draw=magenta!50,fill=magenta!20,thick, minimum width=6mm, minimum height=6mm]
\tikzstyle{nTensor}=[rectangle,draw=blue!50,fill=blue!20,thick,minimum width=20mm,rounded corners=.8ex]
\tikzstyle{colorFill}=[-,fill=magenta!20]
\tikzstyle{trotterTensor} = [rectangle,draw=red!50,fill=red!20,thick, minimum width=16mm, minimum height=6mm]

\tikzstyle{graphnode} = [shape=circle,draw,inner sep=2pt]

\tikzstyle{graphedge} = [->, >=stealth]

\tikzset{
    aten/.style={
        draw,
        fill=MidnightBlue!20,
        inner sep = 0,
        minimum width = 0.6cm,
        minimum height = 0.6cm
}}

\tikzset{
    bten/.style={
        draw,
        fill=OliveGreen!20,
        inner sep = 0,
        minimum width = 0.6cm,
        minimum height = 0.6cm
}}
\tikzset{
    cten/.style={
        draw,
        fill=Dandelion!20,
        inner sep = 0,
        minimum width = 0.6cm,
        minimum height = 0.6cm
}}
\tikzset{
    dten/.style={
        draw,
        fill=Violet!20,
        inner sep = 0,
        minimum width = 0.6cm,
        minimum height = 0.6cm
}}
\tikzset{
    diagten/.style={
        draw,
        fill=Red!20,
        inner sep = 0,
        minimum width = 0.6cm,
        minimum height = 0.6cm
}}

\DeclareMathOperator{\Z}{\mathds{Z}}

\DeclareMathOperator{\U}{\mathrm{U}}
\newcommand{\vdag}{{\phantom\dag}}
\newcommand{\hc}{\mathrm{h.c.}}

\newcommand{\partf}{\mathcal{Z}(\beta)}

\newcommand{\paulx}[1]{\sigma^x_{#1}}
\newcommand{\pauly}[1]{\sigma^y_{#1}}
\newcommand{\paulz}[1]{\sigma^z_{#1}}
\newcommand\norm[1]{\lVert#1\rVert}
\DeclareMathOperator{\Tr}{\mathrm{Tr}}

\newcommand{\hilb}{\mathcal{H}}
\newcommand{\liouv}{\mathcal{L}}

\DeclareMathOperator{\unit}{\mathds{1}}

\newcommand{\der}{\mathrm{d}}

\DeclareMathOperator{\leaszm}{\Gamma_{\epsilon}}
\DeclareMathOperator{\szm}{\Gamma}

\usepackage{standalone}

\begin{document}

\title{Almost Strong Zero Modes at Finite Temperature}

\author{Niklas Tausendpfund}
\affiliation{%
 Forschungszentrum Jülich GmbH, Institute of Quantum Control, Peter Grünberg Institut (PGI-8), 52425 Jülich, Germany
}%
\affiliation{Institute for Theoretical Physics, University of Cologne, D-50937 Cologne, Germany}

\author{Aditi Mitra}
\affiliation{Center for Quantum Phenomena, Department of Physics,
New York University, 726 Broadway, New York, New York, 10003, USA}

\author{Matteo Rizzi}%
\affiliation{%
 Forschungszentrum Jülich GmbH, Institute of Quantum Control, Peter Grünberg Institut (PGI-8), 52425 Jülich, Germany
}%
\affiliation{Institute for Theoretical Physics, University of Cologne, D-50937 Cologne, Germany}
\date{\today}

\begin{abstract}
Interacting fermionic chains exhibit extended regions of topological degeneracy of their ground states as a result of the presence of Majorana or parafermionic zero modes localized at the edges.
In the opposite limit of infinite temperature, the corresponding non-integrable spin chains, obtained via generalized Jordan-Wigner mapping, are known to host so-called Almost Strong Zero Modes, which are long-lived with respect to any bulk excitations.
Here, we study the fairly unexplored territory that bridges these two extreme cases of zero and infinite temperature. 
We blend two established techniques for states, the Lanczos series expansion and a tensor network ansatz, uplifting them to the level of operator algebra.
This allows us to efficiently simulate large system sizes for arbitrarily long timescales and to extract the temperature-dependent decay rates.
We observe that for the Kitaev-Hubbard model, the decay rate of the edge mode depends exponentially on the inverse temperature $\beta$, and on an effective energy scale $\Delta_{\rm eff}$ that is greater than the thermodynamic gap of the system~$\Delta$.

\end{abstract}

\maketitle

\section{\label{sec:intro}Introduction}

Fractionalization of low energy excitations is one of the most interesting properties of topological many-body systems, with the
simplest example of this being the Majorana zero modes ({\bf MZM}s)~\cite{Kitaev2001,Ivanov2001,Nayak2008, Alicea2011,Sarma2015} that appear in fermionic chains protected only by the fermion parity $P$.
These MZMs correspond to fractionalized fermions exponentially localized at the edges of a finite-size system.
Because of this non-local nature, a (topological) ground-state degeneracy arises, as there is no local physical operator that couples to these fractionalized fermions.
The appearance of edge modes can easily be understood in the non-interacting limit as the model formally belongs to the BDI symmetry class in the Altland-Zirnbauer classification~\cite{Altland1997, Ryu2010}. 
For finite interaction strengths, it has been shown that the topological region
-- with the appearance of fractionalized edge modes within the degenerate ground-state manifold --
persists for a wide range of parameters~\cite{Stoundenmire2011, Katsura2015, Jevtic2017, Mahyaeh20}.

Away from the ground-state manifold, little is known about the spectral properties for generic interactions, and whether stable MZMs exist even for excited states. 
However, this is important because it strongly affects the finite temperature lifetime of local edge excitations that have overlap with the MZMs, and has ramifications on practical realization of topologically protected qubits.
In the zero-temperature limit, where only the ground-state manifold contributes to the dynamics, the topological degeneracy leads to an infinite lifetime of these edge excitations.
On the other hand, at infinite temperature, the same edge excitations were shown to have an unusually long lifetime compared to generic bulk excitations~\cite{Kemp2017,Else2017,Scaffidi19,Kemp20,Yates2020,Yates2020_B,Yeh23,Olund2023}.
In integrable limits, this behavior can be explained by the appearance of a Strong Zero Mode ({\bf SZM}) \cite{Kitaev2001,Sen13,Fendley2016,Yates2019,Vernier24}, a generalization of the Majorana zero mode to the full spectrum. Here, the existence of the SZM implies a protected degeneracy of the whole spectrum and not only of the ground-state manifold, and thus one recovers an infinite lifetime.
Away from these special limits, the lifetime becomes strictly finite, with the edge mode often referred to as an Almost Strong Zero Mode ({\bf ASZM}) \cite{Kemp2017}.
Besides some phenomenological arguments \cite{Else2017}, it is still an open question how the lifetime of the edge excitations behaves at finite temperatures. In particular, it is not known how the infinite lifetime emerges as the temperature is lowered from infinity to zero.

In this work, we explore this rather unchartered territory by employing a tensor network ansatz~\cite{Orus2019} to approximate the sequence of operators generated by the Lanczos algorithm for Heisenberg time evolution~\cite{viswanath1994recursion, Parker2019}.
This procedure maps the operator dynamics to the time evolution of a single-particle problem on an artificial one-dimensional chain.
Originally used to study the complexity growth of time-evolved operators, this method has been shown to be a useful tool for understanding the emergence of these long-lived ASZMs~\cite{Yates2020, Yates2020_B}.
However, these earlier studies were highly limited in system size and, most importantly, to infinite temperatures.
The formulation of the Lanczos algorithm for tensor networks removes the limitation of small system sizes by introducing a controlled approximation given by the bond dimension $\chi$ of the tensor network.
Moreover, an intrinsic tensor network formulation allows for the efficient inclusion of arbitrary temperatures $T = 1/\beta$, since the density matrix $\rho(\beta)$ can also be approximated by a tensor network~\cite{Feiguin2005}.
We emphasize that  in contrast to direct integration schemes such as the time-dependent variational principle (TDVP)~\cite{Haegeman2011,Haegeman2016}, the Lanczos method developed here converges quickly with the bond dimension in the case of a long-lived ASZM.
In the former case, the bond dimension necessarily has to grow exponentially with time, while this is not the case for our Lanczos approach, see Appendix \ref{sec:ap:tdvp_vs_lanczos} for details. 

As an application, we explicitly calculate the lifetime of the Majorana edge mode in the non-integrable Kitaev-Hubbard chain~\cite{Maceira2017}.
We find an exponential dependence of the lifetime on the inverse temperature $\tau(\beta) = \exp(\Delta_{\rm{eff}}\beta)$.
However, in contrast to generic bulk excitations where $\Delta_{\rm{eff}}$ is expected to be the energy gap in the many-body spectrum~\cite{Sachdev2011}, we observe a non-trivial dependence of $\Delta_{\rm{eff}}$ on the interaction strength. 
In particular, we find that $\Delta_{\rm{eff}}$ is consistently larger than the many-body gap, hinting at a degenerate structure of the low-energy portion of the spectrum and not exclusively of the ground state.

The paper is organized as follows: In Section~\ref{sec:definitions}, we introduce the main concepts such as the finite temperature autocorrelation function from which the lifetime of an excitation can be extracted, and the Lanczos iteration for calculating the autocorrelation function. We also introduce our algorithm for evaluating the Lanczos series approximately  using tensor networks.
We close this section by reviewing the concept of ASZMs, which are the central object studied in this paper.
In Section~\ref{sec:model_definition}, we introduce the Kitaev-Hubbard chain: a toy model exhibiting an extended topological phase in its ground-state phase-diagram. The Majorana edge modes appearing in this topological phase serve as a perfect test for our algorithm. 
Finally in Section~\ref{sec:results}, we present the numerical findings for the lifetime of these Majorana edge modes at various parameter points in the topological phase of the Kitaev-Hubbard chain. 
We close this paper with Section~\ref{sec:conlusion} that summarizes our findings and comments on possible extensions to other systems such as parafermions~\cite{Fendley2012, Alicea2016, Iemini2017_B}, Floquet circuits~\cite{Sen13, Yates2019,Harper2020, Yates2021,Matthies2022} and number conserving realizations of MZMs~\cite{Cheng2011, Kraus2013, Lang2015, Iemini2017_B,Iemini2017, Lisandrini2022, Tausendpfund2023, Defossez2024, Michen2024}.

\section{\label{sec:definitions}Definitions}

\subsection{\label{sec:acf_ft}Autocorrelation functions at finite temperatures}

We define the lifetime of an excitation generated by the operator $\hat{O}$ by the decay of the autocorrelation function (\textbf{ACF}) defined as
\begin{equation}
    \label{eq:acf_def}
    C_\beta(\hat{O}, t) \coloneqq \braket{\hat{O}|\hat{O}(t)}_\beta\, ,
\end{equation}
with the temperature dependent scalar product \cite{viswanath1994recursion,Parker2019, nandy24}
\begin{equation}
    \label{eq:ft_sclprd}
    \braket{\hat{A}|\hat{O}}_\beta = \frac{1}{2} \Tr\left[
        \rho(\beta) \left\lbrace \hat{A}^\dag \hat{O} + \hat{O}\hat{A}^\dag\right\rbrace
    \right] \, .
\end{equation}
Here $\hat{O}(t) = e^{itH}\hat{O}e^{-itH}$ denotes the Heisenberg time evolution, $\beta = 1/T$ is the inverse temperature, and $\rho(\beta) = \exp(-\beta H)/\Tr\left[\exp(-\beta H)\right]$ is the finite temperature density matrix.

Let us briefly discuss the two limiting cases of Eq.~\eqref{eq:ft_sclprd}, namely of zero and infinite temperature.
In the infinite temperature limit $\beta \to 0$, the scalar product of Eq.~\eqref{eq:ft_sclprd} becomes proportional to the Frobenius scalar product on the vector space of operators
\[
    \braket{\hat{A}|\hat{O}}_{\beta = 0} = \frac{1}{\dim(\hilb)} \Tr[\hat{A}^\dag\hat{O}]\, .
\]
The normalization is given by the dimension of the many-body Hilbert-space $\hilb$.

On the other hand, for zero temperature $\beta \to \infty$, the density matrix projects onto the ground-state manifold, denoted by ${\rm GS}$. 
Thus, the scalar product in Eq.~\eqref{eq:ft_sclprd} reduces to an equally weighted average over all ground-states in ${\rm GS}$:
\[
    \braket{\hat{A}|\hat{O}}_{\beta = \infty} = \frac{1}{2\dim({\rm GS})} \!\sum_{\Omega \in {\rm GS}}\! \braket{\Omega|
            \hat{A}^\dag \hat{O} + \hat{O}\hat{A}^\dag
        |\Omega} \, .
\]

Before closing this subsection, let us comment on the fact that the temperature-dependent scalar product in Eq.~\eqref{eq:ft_sclprd} is not a unique choice, see Appendix \ref{sec:ap:ftsclprd}.
However, our choice appears naturally in linear response theory and directly links the ACF to a measurable quantity~\cite{Parker2019,viswanath1994recursion}.

\subsection{\label{sec:lanzcos_series}Lanczos Series Evaluation of The Autocorrelation Function}

To solve the Heisenberg time evolution, and thus calculate the ACF, we make use of the Lanczos algorithm \cite{viswanath1994recursion}.
As we detail below, this generates a tri-diagonal superoperator that can be interpreted as a single particle hopping on a semi-infinite chain, where the sites are (orthonormal) operators.
In fact, the edge density of states ({\bf EDOS}) of this artificial single particle problem carries all the information about the ACF.

Defining the superoperator $\liouv \hat{O} \coloneqq [H,\hat{O}]$, the Heisenberg time evolution can be written as
\begin{equation}
\label{eq:heisenberg}
\hat{O}(t) = e^{iHt}\hat{O}e^{-iHt} = \sum_{n = 0}^\infty \frac{(it)^n}{n!} \liouv^n\hat{O}\,.
\end{equation}
The Lanczos algorithm now aims at constructing an operator basis $\mathcal{O}_n$ to express the time evolution in a more efficient way.
This basis is constructed to be orthonormal with respect to the temperature dependent scalar product defined in Eq.~\eqref{eq:ft_sclprd}.
To this end, we assume w.l.o.g. $\braket{\hat{O}|\hat{O}}_\beta = 1$ and require $\hat{O}^\dagger = \hat{O}$.

To iteratively construct this new basis, we start by setting $\mathcal{O}_0 \coloneqq \hat{O}$, $\mathcal{O}_{-1} = 0$, and $b_0 = 0$.
The sequence of orthonormal operators then reads
\begin{equation}
\label{eq:lanczos_iteration}
\begin{split}
    \hat{A}_n & = \liouv \mathcal{O}_{n-1} - b_{n-1} \mathcal{O}_{n-2},\\
    b_n &= \sqrt{\braket{\hat{A}_n|\hat{A}_n}_\beta}, \\
    \mathcal{O}_n & = \hat{A}_n / b_n \, .
\end{split}
\end{equation}
The time evolved operator $\hat{O}(t)$ can be expanded in this basis with real coefficients $\varphi_n(t)$ as
\[
\hat{O}(t) = \sum_{n=0}^\infty i^n \varphi_n(t) \mathcal{O}_n\,,\ \varphi_n(0) = \delta_{n,0} \, .
\]
Further, by defining the states $\ket{n} = i^n \mathcal{O}_n$, with $\braket{n|m} = \delta_{m,n}$, the ACF is now equivalently expressed by
\begin{align}
\label{eq:acf_artf_spham}
    C_\beta(\hat{O},t) = \varphi_0(t) = \braket{0|e^{-itH_{\rm sp}}|0} \, , \\
\label{eq:artf_sp_ham}
    H_{\rm sp} = \sum_{n=0}^\infty i\, b_{n+1}\ket{n+1}\bra{n} + \hc \, ,
\end{align}
where $H_{\rm sp}$ is the artificial single particle Hamiltonian, solely defined by the Lanczos coefficients $b_n$%
\footnote{We underline that, for a chain with open boundaries, complex phases of hopping coefficients do not play any role.}.
In fact, these Lanczos coefficients $b_n$ carry all the information about the seed operator $\hat{O}$ used to construct the Lanczos series, the Hamiltonian $H$, and the temperature through the chosen scalar product.

Instead of computing the ACF directly by calculating the exponential of Eq.~\eqref{eq:artf_sp_ham}, it is simpler to reconstruct the ACF from the EDOS of the single particle Hamiltonian $H_{\rm sp}$, defined as
\begin{equation}
    \label{eq:edge_dos}
    \nu_\beta^E(\omega) = \braket{0|\delta(\omega - H_{\rm sp})|0}\, ,
\end{equation}
The EDOS $\nu_\beta^E(\omega)$ is connected to the ACF $C_\beta(\hat{O},t)$ by a simple Fourier transform
\begin{equation}\label{eq:acf_dos_fourier}
    C_\beta(\hat{O},t) = \int_{-\infty}^\infty\! \der\omega \,\nu_\beta^E(\omega) \cos(\omega t),
\end{equation}
and equivalently carries all information about the dynamics.
Note that the tridiagonal structure of $H_{\rm sp}$, Eq.~\eqref{eq:artf_sp_ham}, with zeros on the diagonal, implies $\nu_\beta^E(\omega) = \nu_\beta^E(-\omega)$.

In principle, to obtain the full time dynamics of a given operator $\hat{O}$, it is necessary to calculate a large number of the Lanczos coefficients $b_n$.
To keep the computational effort to a minimum, we need to truncate the series at some point.
This is possible because the Lanczos coefficients are expected to grow nearly linearly with $n$, saturating at some plateau value due to finite size effects~\cite{Parker2019}.
We therefore adopt a strategy similar to that used by one of us in Ref.~\cite{Yates2020}.
For this, we compute the first $N$ coefficients of the Lanczos series explicitly. 
After reaching the plateau value, we approximate the unknown values for $n>N$ by setting $b_{n>N} = b_N$.
In terms of the artificial single particle Hamiltonian, this amounts to attaching a featureless semi-infinite homogeneous chain with a hopping parameter $w = ib_N$. 
This approximation also leads to an efficient calculation of the EDOS in terms of a continued fraction as explained in Appendix \ref{sec:ap:acf_greensfunctions_cntfrct}.

Note that the details of the transition to the semi-infinite chain with uniform hopping is not that crucial, see also Appendix \ref{sec:ap:convergence} for a numerical demonstration. 
This can be understood in terms of the artificial Hamiltonian $H_{\rm sp}$. 
Since we are mainly interested in the edge properties of $H_{\rm sp}$, changing the parameters far away from the edge has only a small influence, as long as $N$ is large enough and $b_N$ is placed well inside the featureless plateau.

\subsection{\label{sec:tn_lanczos}Evaluation using a Tensor Network Ansatz}

The tridiagonal form of the artificial $H_{\rm sp}$ of Eq.~\eqref{eq:artf_sp_ham} does not imply that the exact calculation of the Lanczos series is an easy task overall. 
Indeed, since the Lanczos iteration involves nested commutators, the basis states $\mathcal{O}_n$ quickly become fully dense matrices for any given system size.
%
The exponential growth $2^{2L}$ with system size of the number of elements, limits previous studies~\cite{Yates2020, Yates2020_B} to very small system sizes and infinite temperatures.
Including finite temperatures would require a full diagonalization of the Hamiltonian to obtain the density matrix $\rho(\beta)$, a task impossible for $L>16$.
To overcome this limitation in system size and temperature, we introduce a tensor network approach that approximates the basis operators $\mathcal{O}_n$.
%
%
%
In particular, we choose the operators $\mathcal{O}_n$ to be represented by a matrix product operator (\textbf{MPO}):

\vspace{2ex}

\begin{center}
\begin{tikzpicture}[baseline, scale = 0.85]
        \node[anchor = east] at (0.0,0) {$\mathcal{O}_n = $};
        
        \node[aten] (a1) at (0.5, 0.) {$O^n_1$};
        \node[aten] (a2) at (2, 0.) {$O^n_2$};
        \node[aten] (a3) at (3.5, 0.) {$O^n_3$};

        \draw (a1) -- (a2) node [midway, above] {$\chi$};
        \draw (a2) -- (a3) node [midway, above] {$\chi$};;
        \draw (a3) -- ($(a3) + (1, 0)$) node [midway, above] {$\chi$};

        \node at ($(a1) + (0, 0.9)$) {$\sigma_{1}$};
        \node at ($(a2) + (0, 0.9)$) {$\sigma_{2}$};
        \node at ($(a3) + (0, 0.9)$) {$\sigma_{3}$};
        
        \draw (a1) -- ($(a1) + (0,0.6)$);
        \draw (a2) -- ($(a2) + (0,0.6)$);
        \draw (a3) -- ($(a3) + (0,0.6)$);
        
        \node at ($(a1) - (0, 0.9)$) {$\sigma^\prime_{1}$};
        \node at ($(a2) - (0, 0.9)$) {$\sigma^\prime_{2}$};
        \node at ($(a3) - (0, 0.9)$) {$\sigma^\prime_{3}$};
        
        \draw (a1) -- ($(a1) - (0,0.6)$);
        \draw (a2) -- ($(a2) - (0,0.6)$);
        \draw (a3) -- ($(a3) - (0,0.6)$);
        
        \node[anchor=west] at (4.5, 0) {$\dots\quad .$};

\end{tikzpicture}
\end{center}

\vspace{2ex}

Here $\chi$ is the bond dimension of the ansatz and is kept fixed.
For a given bond dimension $\chi$, the number of elements grows algebraically with the system size $\mathcal{O}(L\chi^2d^2)$, in contrast to the exponential growth of dense matrices.
Similarly, the Hamiltonian $H$ and the density matrix $\rho(\beta)$ can also be efficiently represented by an MPO~\cite{Verstraete2004,Feiguin2005}.
To evaluate the Lanczos iteration of Eq.~\eqref{eq:lanczos_iteration}, we have to replace the normal matrix algebra by the corresponding tensor network algebra.
For example, the application of the superoperator $\liouv\mathcal{O}_n = H\mathcal{O}_n - \mathcal{O}_nH$ can be expressed by two MPO-MPO applications followed by subtraction of the two resulting MPOs.
In general, these tensor network operations lead to an increase in the bond dimension of the final MPO.
To keep the bond dimension fixed at $\chi$, it is therefore necessary to truncate the bond dimension back to the target bond dimension $\chi$.

In this paper we choose to apply a truncation scheme based on the singular values of the MPO.
More specifically, we use the reduced density matrix approach~\cite{McCulloch2007}, analogously to matrix product states.
Moreover, to make the application of the superoperator as exact as possible, we apply this truncation scheme in one shot to the full $\liouv \mathcal{O}_n$ and not separately after each intermediate step (i.e., $H\mathcal{O}_n$, $\mathcal{O}_nH$ and their subtraction).
Let us note that this truncation scheme maximizes the fidelity $\braket{\hat{M}|\liouv\mathcal{O}_n}_0$, which corresponds to the infinite-temperature scalar product. 
In principle, one could also directly optimize the fidelity obtained from the finite-temperature scalar product via a variational optimization.

\begin{algorithm}[H]
\caption{Lanczos Iteration}
\label{alg:lanczos_step}
\begin{algorithmic}
\Function {lanczos\_step}{$\mathcal{O}_n$, $\mathcal{O}_{n-1}$, $b_n$, $H$, $\beta$, $\chi$}
    \State $\hat{B}_{n+1} \gets \texttt{commutator}(H,\, O_n,\, \chi)$
    \State $\hat{A}_{n+1} \gets \texttt{add}(\hat{B}_{n+1},\, -b_n \cdot  \mathcal{O}_{n-1},\, \chi)$
    \State $b_{n+1} \gets \texttt{norm}(\hat{A}_{n+1},\, \beta)$
    \State $\mathcal{O}_{n+1} \gets \hat{A}_{n+1}/b_{n+1}$
    \State \Return $(\mathcal{O}_{n+1}, b_{n+1})$
\EndFunction
\end{algorithmic}
\end{algorithm}

The complete Lanczos iteration step described by Eq.~\eqref{eq:lanczos_iteration} using the tensor network approach is summarized in the pseudo-code~\ref{alg:lanczos_step}.
The core function \textrm{LANCZOS\_STEP} takes as input the current $\mathcal{O}_n$ and the previous basis operator $\mathcal{O}_{n-1}$, the current Lanczos coefficient $b_n$, the system Hamiltonian $H$ and the inverse temperature $\beta$ together with the target bond dimension $\chi$.
The function \texttt{commutator} calculates the application of $\liouv$ to $\mathcal{O}_n$ in MPO form with a direct truncation of the target bond dimension to $\chi$, see Appendix \ref{sec:ap:implementation_details}. 
Similarly, \texttt{add} takes the output of \texttt{commutator}, subtracts $b_n\mathcal{O}_{n-1}$ as a MPO with a direct truncation and stores the result in $\hat{A}_{n+1}$.
Finally \texttt{norm} calculates the norm of $\hat{A}_{n+1}$ with respect to the temperature dependent scalar product of Eq.~\eqref{eq:ft_sclprd}.

Note that the algorithm presented here can be directly applied in the thermodynamic limit, at least for infinite temperature where $\rho = \unit$ independent of the system size.
For example, for a seed operator $\mathcal{O}$ originally localized at the edge of a semi-infinite chain and a Hamiltonian $H$ with only nearest neighbor interaction, the basis operators $\mathcal{O}_n$ have a support on at most $n+1$ sites.
Thus, by growing the operator in each step, it is possible to remove all finite size effects.
For finite temperatures one needs a good representation of the density matrix $\rho(\beta)$ for an infinite system, which is beyond the scope of this article.


\subsection{\label{sec:aszm_intro}(Almost) Strong Zero Modes}

MZMs occur in gapped fermionic chains where the number conservation is broken to a $\Z_2$ fermionic parity symmetry.
They always appear in pairs, with one MZM at the left and the other at the right end of the chain.
The appearance of a MZM is associated with a ground-state degeneracy of two, where the two ground-states $\ket{\Omega_p}$ have opposite parity $p=\pm$. 
As they only differ in the occupancy of the MZM, there is no local bulk operator that can distinguish between the two ground-states and the degeneracy is topologically protected.
To define the MZM, we assume that the chain is populated by spinless fermions described by the creation/annihilation operators $c_j^\dag/c_j^\vdag$.
Equivalently, we can define the set of Majorana operators by $c_j = (\gamma_{j,a} - i\gamma_{j,b})/2$.

Now, given the two ground-states, one typically defines the MZM as
\begin{equation}
\begin{split}
    \gamma_L \coloneqq \sum_{j=1}^N \varphi_j \gamma_{j,a} \,,\ 
    \varphi_j \coloneqq \Re\left[\braket{\Omega_+|\gamma_{j,\alpha}|\Omega_-}\right] \,.
\end{split}
\end{equation}
Above, $\Re[z]$ denotes the real part of the complex number $z$.
Here for simplicity, we focus on the left end of the chain and assume that only the $\gamma_{j,a}$ Majorana operators contribute (in contrast to odd products of the Majoranas).
A typical behavior is exponential localization with $|\varphi_j| \sim e^{-\delta j}$, with $\delta$ depending on the coupling parameters of the model \cite{Kitaev2001}.

The existence of a MZM has strong consequences for the ACF for $\gamma_{1,a}$ at zero temperature.
As discussed in subsection~\ref{sec:acf_ft}, the ACF reduces to an equally weighted average over correlation functions in the ground-state manifold, i.e.
\[
C_{\beta =\infty}(\gamma_{1,a}, t) = \frac{1}{2}\sum_{p = \pm 1}\!\! \Re\left[\braket{\Omega_{p}|\gamma_{1,a}e^{iHt}\gamma_{1,a}|\Omega_{p}}\right],
\]
where we have assumed $H\ket{\Omega_P} = 0$.
From $\bar{p} = -p$, it follows from an insertion of the identity that:
\[
\begin{split}
    &\Re\left[\braket{\Omega_{p}|\gamma_{1,a}e^{iHt}\gamma_{1,a}|\Omega_{p}}\right] = 
    \\ &{}|\braket{\Omega_{p}|\gamma_{1,a}|\Omega_{\bar{p}}}|^2
   \!+\!\! \sum_{n\notin {\rm GS}} |\braket{\Omega_{p}|\gamma_{1,a}|n, \bar{p}}|^2\! \cos(E_{n,\bar{P}}t)\\
   {}&= |\varphi_{1}|^2 + \tilde{C}(t) \xrightarrow[t\to\infty]{} |\varphi_{1}|^2 \, ,
\end{split}
\]
where $\tilde{C}(t)$ represents the incoherent part coming from the states above the gap and is assumed to decay rapidly.
It follows $C_\infty(\gamma_{1,a},t) \to |\varphi_{1}|^2$ for $t\to\infty$.

In the previous discussion, the MZM was defined solely by the properties of the ground-state manifold.
The SZM can be seen as a generalization of these ideas to the full many-body spectrum.
In this perspective, a SZM~\cite{Kitaev2001,Fendley2012, Jermyn2014, Fendley2016,Alicea2016} is defined as an operator $\szm$ with the following properties
\begin{itemize}
    \item[1.] Hermitian: $\szm^\dag = \szm$,
    \item[2.] Anti-commuting with the fermionic parity: $\lbrace P, \szm \rbrace = 0$,
    \item[3.] Commuting with the Hamiltonian: $[\szm, H]\to 0$ for $L\to \infty$.
\end{itemize}
We also require the $\szm$ to be localized at the edge of the system, so that the SZM has an exponentially decaying weight on operators with support away from the edge of the chain, similar to the MZM.

It follows that a system possessing a SZM has an exact double degeneracy of the spectrum in the thermodynamic limit:
Every energy eigenstate of defined parity $p$ has a partner state of the opposite parity $-p$ \cite{Kitaev2001,Fendley2012, Jermyn2014, Alicea2016,Fendley2016}.
As a direct consequence, it is easy to show that for any operator with $\braket{\hat{O}, \szm}_\beta = \alpha$ the long-time thermodynamic behavior of the ACF is given by $C_{\beta}(\hat{O},t) \to |\alpha|^2$ for arbitrary temperatures $T = 1/\beta$.
In contrast, the existence of a MZM only guarantees the double degeneracy in the ground-state manifold and thus an infinite lifetime of $\gamma_{1,a}$ only at exactly zero temperature.

The construction of a SZM in an exact way has only been achieved in a few cases of integrable Hamiltonian models~\cite{Kitaev2001, Fendley2016}.
Apart from integrability, it is still an open question whether it is possible to find a SZM and whether it is possible to have an exact double degeneracy in the spectrum even in a perturbative regime~\cite{Kells15,Kells15_B,Mahyaeh20}.
Still, it has been found numerically~\cite{Else2017, Kemp2017, Scaffidi19,Kemp20,Yates2020,Yates2020_B} that $\gamma_{1,a}$ has a long lifetime at infinite temperature. 
Unlike a SZM, in this case the lifetime saturates with the system size and is strictly finite in the thermodynamic limit.
This behavior has been linked to the existence of an ASZM.  
More precisely, an ASZM shares all properties of a SZM except that the commutator with the Hamiltonian saturates to a non-zero operator with increasing system size~\cite{Else2017}.
This error term then necessarily leads to a finite lifetime of the edge excitation $\gamma_{1,a}$.

In a pre-thermal regime, the authors of~\cite{Else2017} connected the appearance of such a nearly commuting operator to an approximately conserved $U(1)$ symmetry. 
In their formulation, the ASZM is given by a local unitary rotation of $\gamma_{1,a}$.
A different approach to understanding the ASZM was proposed by Yates et al.~\cite{Yates2020,Yates2020_B}, who linked this behavior to the Lanczos series $b_n$ obtained by using $\gamma_{1,a}$ as the seed operator.
In particular they found that the artificial single particle Hamiltonian $H_{\rm sp}$ from Eq.~\eqref{eq:artf_sp_ham} resembles that of a dressed Su-Schrieffer-Heeger\cite{Su1979,Su1980} (SSH) model with a vanishing staggering:
\begin{equation}\label{eq:stag_Lanczos}
    b_n = h_n + (-1)^n \tilde{h}_n\, .
\end{equation}
Here $h_n$ is the positive monotonically increasing background hopping in Krylov subspace, and is expected to be present for any generic chaotic models~\cite{Parker2019}, while $\tilde{h}_n$ is the staggered component, which becomes trivial for some $n > n^\star$.
This structure has a strong influence on the possible form of the EDOS $\nu_\beta^E(\omega)$, Eq.~\eqref{eq:edge_dos}.

Here, we make use of the fact that the EDOS must to be of the form
\begin{equation}
\label{eq:dos_general_form}
    \begin{split}
        \nu_\beta^E(\omega) = &A(\beta) \frac{\gamma(\beta)/\pi}{\omega^2 + \gamma(\beta)^2} \\
        {}&+ \left[1 - A(\beta)\right]\tilde{\nu}_\beta^E(\omega)\, ,
    \end{split}
\end{equation}
where $\tilde{\nu}_\beta^E(\omega)$ defines an incoherent background density of states with a gap around zero energy.
This incoherent background leads to fast short-time dynamics, while the asymptotic behavior is dominated by the Lorentzian line-shape of width $\gamma(\beta)$.

In fact, the general model of Eq.~\eqref{eq:stag_Lanczos} can be mapped by simple arguments to a new model consisting of a SSH chain of length $N_{\rm eff}$ attached to a semi-infinite lead with homogeneous hopping. 
The Lorentzian peak results from the hybridization of the topological edge state of the SSH model with the gapless spectrum of the semi-infinite lead, while the bulk modes the SSH chain give rise to sidebands approximately described by semicircles:
\begin{figure*}[ht!]
     \centering
    \includegraphics[width=0.8\linewidth]{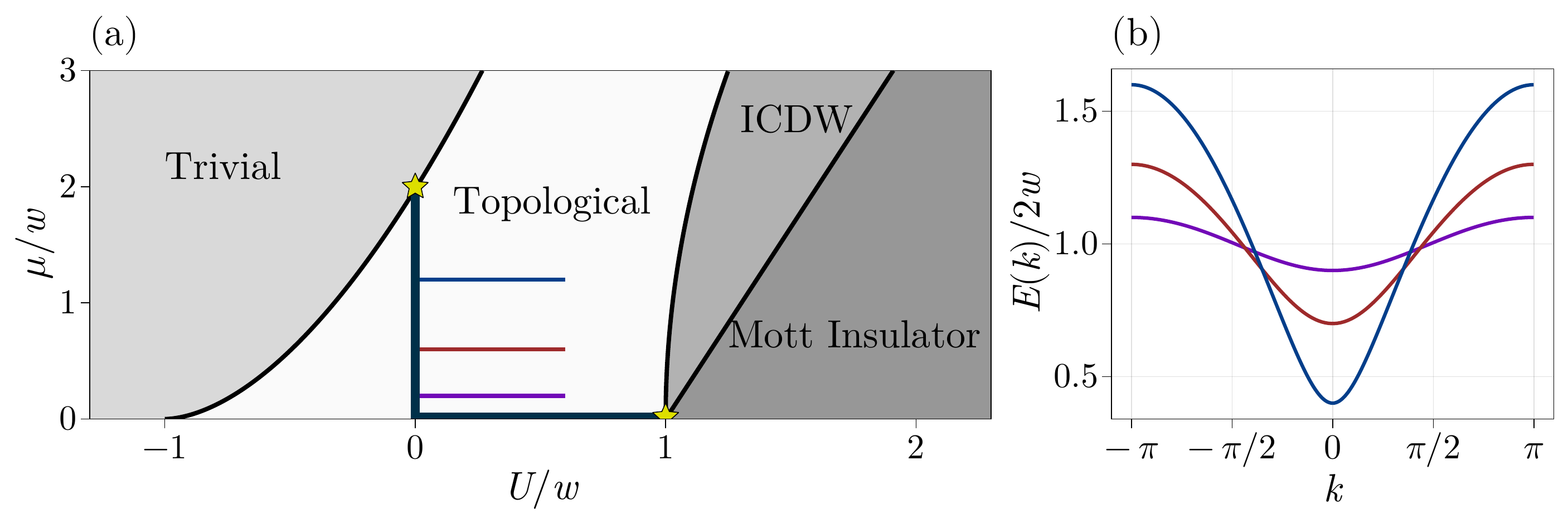}
    \caption{
    (a) Sketch of the different phases of the Hamiltonian in Eq.~\eqref{eq:kitaev_hubbard_chain} following \cite{Hassler2012, Katsura2015, Mahyaeh20}.
    For small $\mu/w$ and $U/w$, there exists an extended region (white) showing a topological ground-state degeneracy.
    Increasing either $\mu$ or $U$, one observes phase transitions into topologically trivial regions.
    For large $\mu$, the state is described by a trivial band insulator, while for large $U$, one observes a transition into an incommensurate charge density wave phase, followed by a commensurate-incommensurate transition into a Mott insulating phase.
    The black lines denote the exactly solvable limits where it is possible to construct SZMs that commute with the Hamiltonian. These SZMs disappear at the phase transitions to the trivial regions, marked by yellow stars.
    Along the three cuts, marked by purple ($\mu/w = 0.2$), red ($\mu/w = 0.6$), and blue ($\mu/w = 1.2$), we compute the effective energy scale $\Delta_{\rm eff}$ in Sec.~\ref{sec:results:effective_gaps}.
    (b) Energy dispersion relations for $U=0$ are displayed and with the identification $E(k=0)=\Delta(U=0)$. 
    }
    \label{fig:phasediagram_sketch}
\end{figure*}
\begin{equation}
\label{eq:sidebands_dos_mdl}
    \begin{split}
        2\tilde{\nu}_\beta^E(\omega) &= \nu_{\rm C}(E - E_0) + \nu_{\rm C}(E + E_0),\\
        \nu_{\rm C}(\omega) &= \frac{1}{\pi w^\star}\sqrt{1 - \frac{\omega^2}{(2w^\star)^2}}\ \theta(2w^\star - |\omega|) \, .
    \end{split}
\end{equation}
See Appendix \ref{sec:app:edos_ssh_connection} for more details.
We propose a simple model to capture all the dynamics of the ASZM at short and long times, involving four fitting parameters $(A,\gamma, E_0, w^\star)$.
In this model, the appearance of a narrow Lorentzian peak is the signature of an ASZM.
From the EDOS, one can also recover the SZM limit as follows: since the Lorentzian contributes to the ACF as $e^{-\gamma(\beta)t}$, an infinite lifetime is recovered only for $\gamma(\beta) \to 0$.
For this case, the Lorentzian function reduces to a delta function $\delta(\omega)$.


\section{\label{sec:model_definition}Model}

The explicit model studied in this work is that of spinless fermions defined by the creation/annihilation operators $c_j^\dag$/$c_j^\vdag$ that reside on a chain of length $L$, and interact according to the 
Kitaev-Hubbard Hamiltonian
\begin{equation}
    \label{eq:kitaev_hubbard_chain}
    \begin{split}
         H = -w\sum_{j = 1}^{L-1} &\, \left(c_j^\dag - c_j^\vdag \right)
         \left(c_{j+1}^\dag + c_{j+1}^\vdag\right) \\
   {} &+ U\, \sum_{j = 1}^{L-1} p_j p_{j+1} - \frac{\mu}{2}\,\sum_{j = 1}^L p_j \, .
    \end{split}
\end{equation}
Above, $p_j = 2c_j^\dag c_j^\vdag - 1$ defines the local parity of the site $j$.
The quadratic part of this Hamiltonian ($U=0$) consists of the usual nearest-neighbor hopping term, a $p$-wave pair creation/annihilation process of neighboring particles, and a chemical potential $\mu$ controlling the average density.
For simplicity, we choose the pairing potential to be equal to the hopping amplitude and denote it by $w$.
In order to break integrability, we introduce a nearest-neighbor Hubbard-like interaction of strength $U$.
The Hamiltonian Eq.~\eqref{eq:kitaev_hubbard_chain} commutes with the total fermionic parity $P =\prod_j p_j$, thus splitting the spectrum into two towers of even and odd parity.

In terms of the Majorana operators introduced in section~\ref{sec:aszm_intro}, the Hamiltonian \eqref{eq:kitaev_hubbard_chain} assumes the form
\begin{equation*}
    \begin{split}
        H = -&w\sum_{j=1}^{L-1} i\gamma_{j,b}\gamma_{j+1,a} +\frac{\mu}{2}\sum_{j = 1}^L i\gamma_{j,a} \gamma_{j,b} \\
        {} - &U\, \sum_{j=1}^{L-1}\gamma_{j,a}\gamma_{j,b}\gamma_{j+1, a}\gamma_{j+1,b} \, .
    \end{split}
\end{equation*}

The Kitaev-Hubbard chain \eqref{eq:kitaev_hubbard_chain} possesses a rich phase-diagram, sketched in Fig.~\ref{fig:phasediagram_sketch}, including an extended topological phase~\cite{Hassler2012, Katsura2015, Mahyaeh20} characterized by a doubly degenerate ground-state manifold $\ket{\Omega_p}$ with opposite fermion parity $p$, together with the appearance of edge-localized MZMs.
While non-integrable for a generic choice of parameters, there are two exactly solvable limits. 
The first limit is the non-interacting case with $U=0$ where the model becomes quadratic in terms of the Majorana operators and is equivalent to the Kitaev chain~\cite{Kitaev2001}.
The second limit corresponds to $\mu=0$ but arbitrary interaction strengths $U$.
In this case, the model is diagonalizable by defining a non-local unitary transformation of the original fermions $c_j$, see Appendix \ref{sec:ap:jw_sec_line} for more information.

In both cases one can construct a SZM analytically~\cite{Kitaev2001, Fendley2016}.
Away from these integrable limits, this is no longer possible.
Nevertheless, the topological ground-state degeneracy does still allow for the existence of a MZM of the form $\gamma_L = \sum_{j=1}^N \varphi_j \gamma_{j,a}$, see discussion in section~\ref{sec:aszm_intro}.


\section{\label{sec:results}Results}

In this section, we discuss the numerical results obtained for the temperature dependent Lanczos series.
Motivated by the analytical results for the SZM in the integrable limits of the Kitaev-Hubbard chain, Eq.~\eqref{eq:kitaev_hubbard_chain}, and the form of the MZM in the ground-state manifold, we choose the edge Majorana operator $\gamma_{1,a} = c_1^\dag + c_1^\vdag$ as the seed operator for the Lanczos algorithm.

\subsection{General behavior of the Lanczos series}

We will start by considering the exemplary point $\mu/w = 1.2$, $U/w=0.1$ to discuss the general features observed at finite temperatures.
For all results we have chosen a fixed system size of $L=22$. 
We have checked that the resulting ACF has converged with respect to the system size, see also Appendix~\ref{sec:ap:convergence}
\begin{figure}[ht!]
    \centering
    \includegraphics[width=\linewidth]{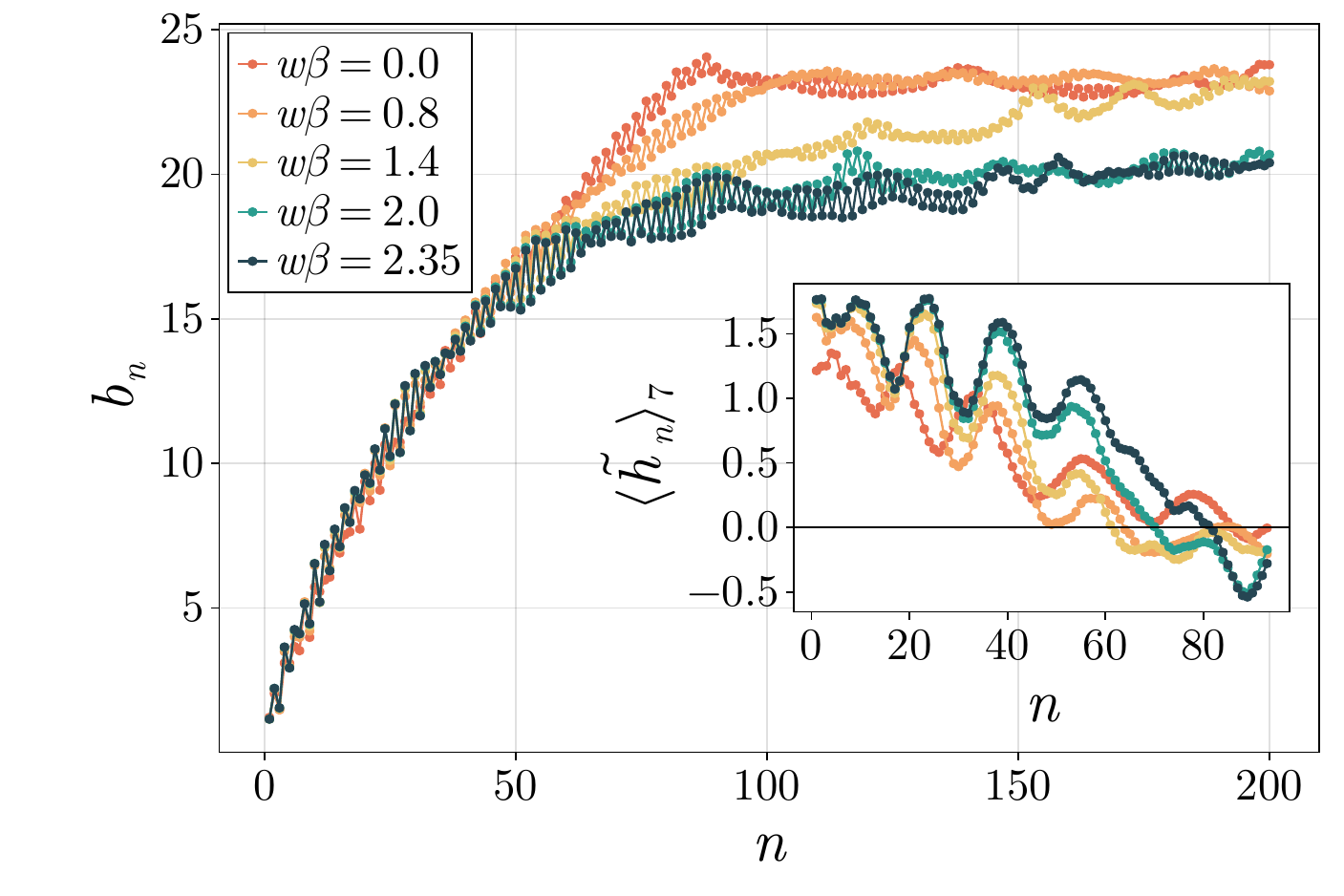}
    \caption{
    Lanczos coefficients for $L=22$ at various inverse temperatures $w\beta = 1/T$. 
    The insert shows the staggered component $\tilde{h}_n$, Eq.~\eqref{eq:staggered_component}, averaged over seven sites to reduce the noise in the data.}
    \label{fig:lanczos_temp}
\end{figure}
We start by discussing the general behavior of the temperature dependent Lanczos series.
The results are obtained by using the tensor network ansatz introduced in Sec.~\ref{sec:tn_lanczos} with a maximal bond dimension of $\chi = 2000$ for the matrix product operator.
See Appendix \ref{sec:ap:convergence} for a detailed discussion on the convergence properties with the bond dimension.
In Fig.~\ref{fig:lanczos_temp} we show the coefficients $b_n(\beta)$ for different inverse temperatures $\beta = 1/T$.
As a generic feature, we observe an increase of the coefficients with respect to $n$ independent of~$\beta$. 

\begin{figure}[ht!]
    \centering
    \includegraphics[width=\linewidth]{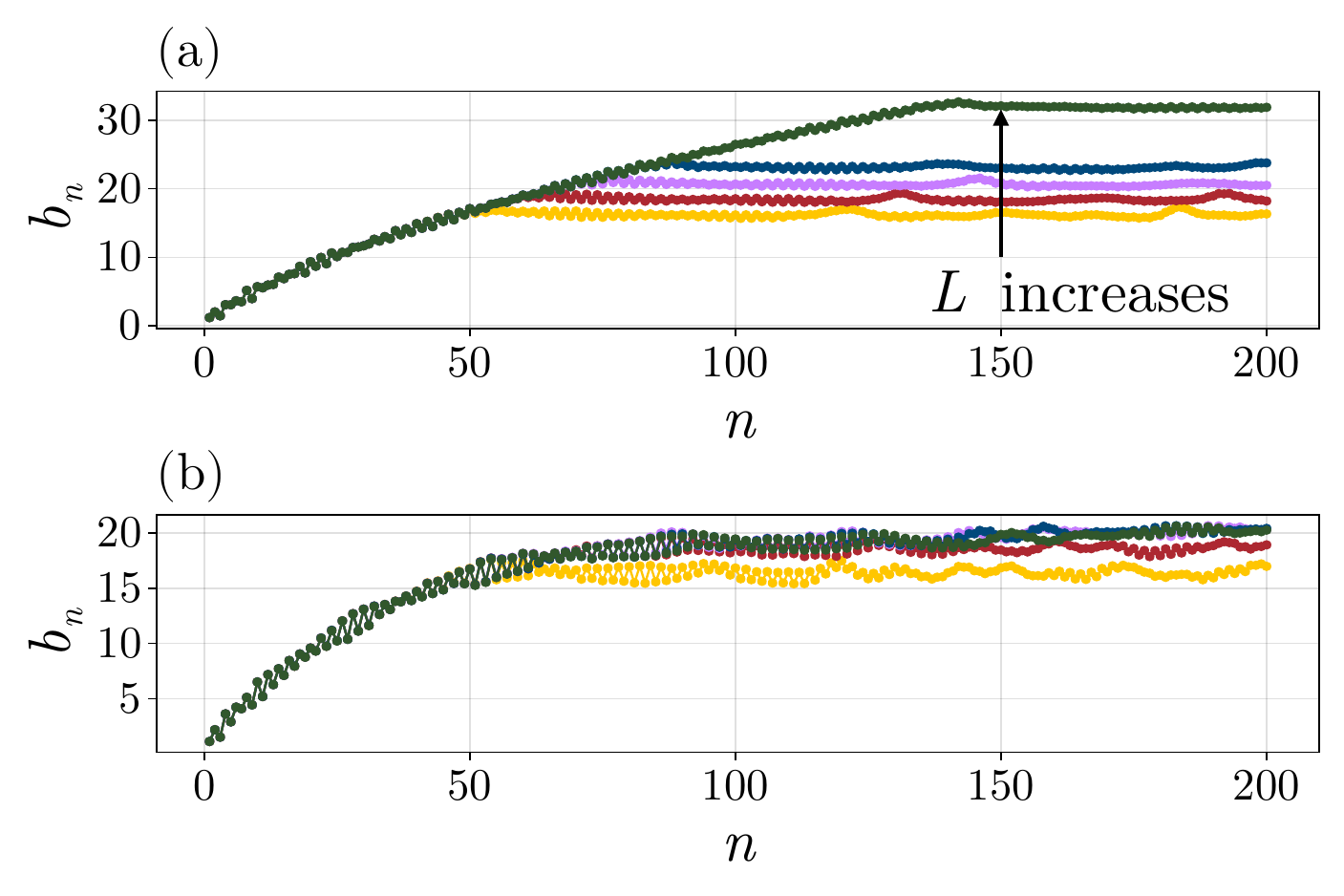}
    \caption{
            (a) Comparison of the Lanczos series with increasing system sizes $L = 16,18,20,22,30$ at infinite temperature. The final plateau value increases with system size.
            (b) The Lanczos sequence for the same $L$ and for $w\beta = 2.35$. In contrast to the infinite temperature sequence, the plateau value depends only weakly on the system size, with almost no difference between $L=22$ and $L=30$.
    }
    \label{fig:lanczos_systemsize_dep}
\end{figure}

For small $\beta$, the increase follows a near linear behavior $b_n \sim n$ as expected for generic non-integrable systems~\cite{Parker2019}, before saturating to a plateau which depends on the system size, see Fig.~\ref{fig:lanczos_systemsize_dep}(a).
For larger $\beta$, the increase starts to deviate from this near linearity with a slightly stronger curvature, more like a power law behavior $n^{\alpha}$, a deeper analysis is left for future work.
The system size dependence is greatly reduced at lower temperatures, as can be seen in Fig.~\ref{fig:lanczos_systemsize_dep}(b). This is expected because as the temperature is lowered, the dynamics is projected to smaller regions of the Hilbert space.

In addition to this general increase, the series is dressed by a staggered component
\begin{equation}
\label{eq:staggered_component}
    \tilde{h}_n \coloneqq (-1)^{n+1}(b_{n+1} - b_n) \,.
\end{equation}
As can be seen by the inset of Fig.~\ref{fig:lanczos_temp}, for all temperatures this staggered component becomes trivial (i.e, either negative or oscillates around zero) for $n>n^*$.
Increasing $\beta$ has the effect of increasing $\tilde{h}_n$, while also  shifting the point $n^*$ at which $\tilde{h}_n \approx 0$, to larger values of $n$. 
The associated artificial single particle Hamiltonian $H_{\rm sp}$ is that of a dressed SSH chain in the topological regime with a vanishing staggering, see also the discussion in subsection~\ref{sec:aszm_intro}. 
We expect that at any finite temperature, the increase in the background, $h_n$, to have only a small influence on the lifetime of $\gamma_{1,a}$ in contrast to the staggered component $\tilde{h}_n$.

\begin{figure}[ht!]
    \centering
    \includegraphics[width=\linewidth]{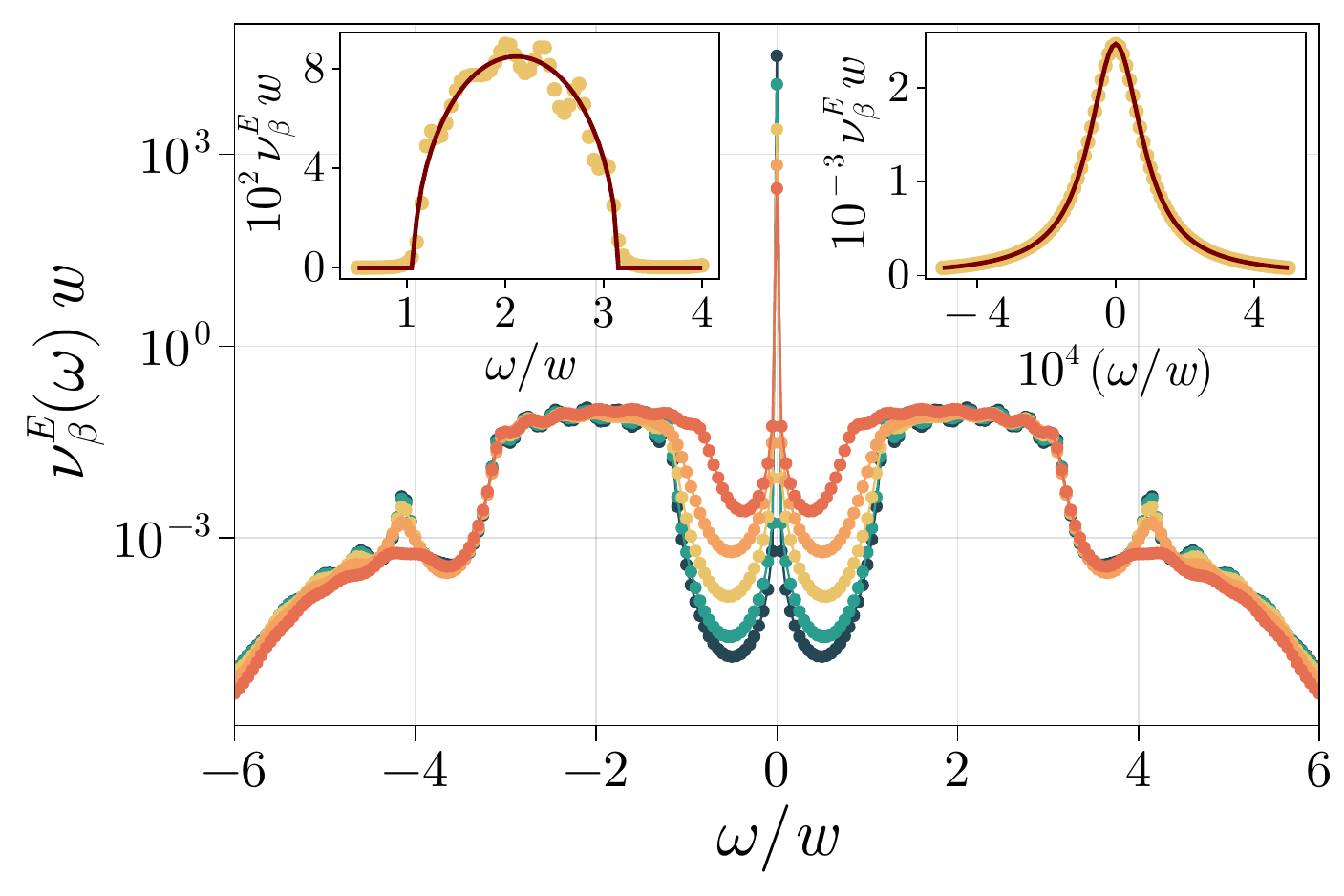}
    \caption{
    EDOS obtained from the Lanczos coefficients at the finite temperatures shown in Fig.~\ref{fig:lanczos_temp}, and with the same color codes. 
    %
    %
    The left inset shows a fit of the sidebands by the simplified model of Eq.~\eqref{eq:sidebands_dos_mdl}. The right inset shows a fit of the central Lorentzian peak. Both insets are for $w\beta = 1.4$.}
    \label{fig:dos_temp}
\end{figure}

This can be made more rigorous by considering the EDOS shown in Fig.~\ref{fig:dos_temp}.
For all temperatures, the general shape of the EDOS is given by a narrow Lorentzian peak around $\omega = 0$, with an additional incoherent background, see Eq.~\eqref{eq:dos_general_form}.
Lowering the temperature has two effects: First, the Lorentzian peak becomes narrower, i.e. the width parameter $\gamma(\beta)$ becomes smaller for larger $\beta$. 
Secondly, the incoherent part changes its form slightly.
While the incoherent background shows a two-band structure separated by an energy gap for all temperatures, additional local extrema appear at higher energies, on lowering the temperature.
We believe that the changes to the incoherent part of the DOS is mainly influenced by the changes to the background values of the Lanczos coefficients $h_n$, while the decrease of $\gamma(\beta)$ is directly related to the increase in the staggered component $\tilde{h}_n$.

\begin{figure}[htbp!]
    \centering
    \includegraphics[width=\linewidth]{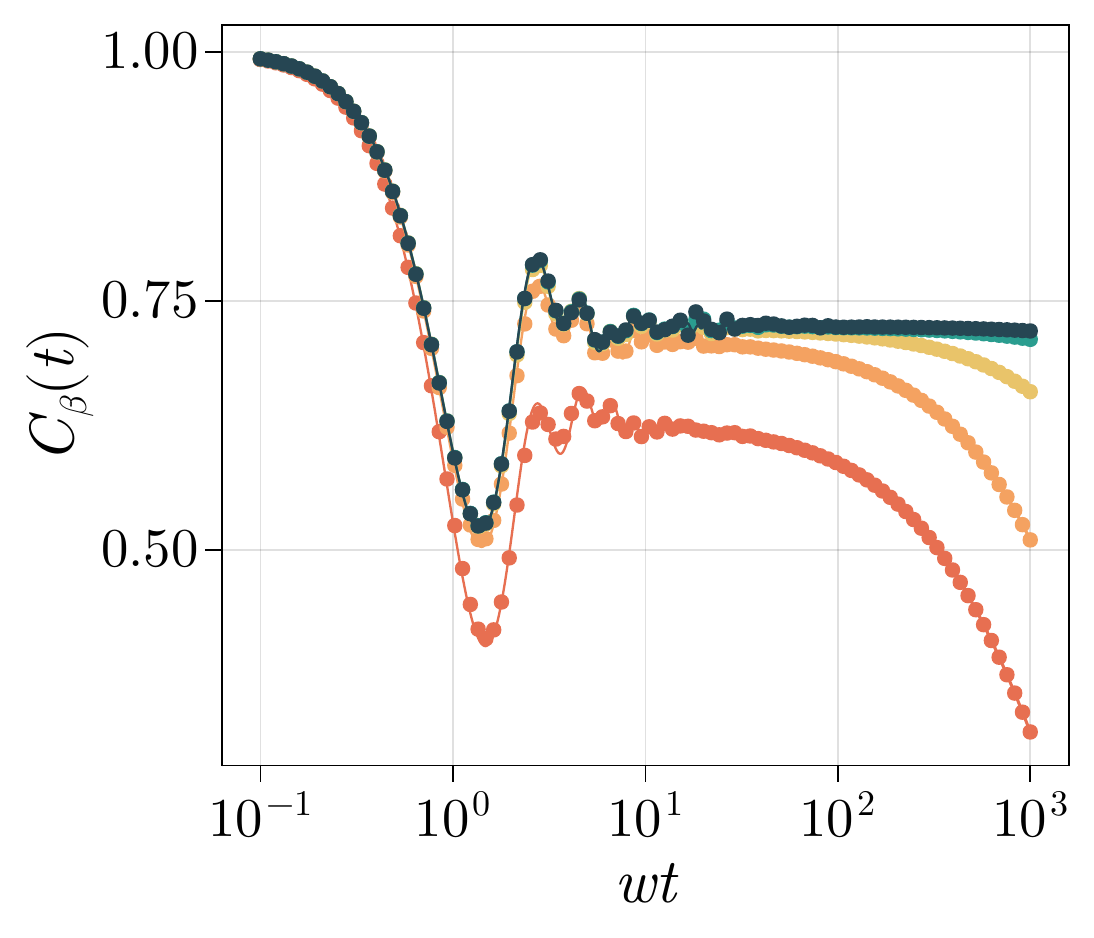}
    \caption{
    ACR for different temperatures with the same color code as in Fig.~\ref{fig:lanczos_temp}.
    The dots represent the ACF obtained from the EDOS from Fig.~\ref{fig:dos_temp}. The lines are the analytical ACF from Eq.~\eqref{eq:acf_simplified_model} with parameters obtained by fitting the EDOS with the simplified model 
    Eq.~\eqref{eq:dos_general_form} and Eq.~\eqref{eq:sidebands_dos_mdl}.}
    \label{fig:acf_temp}
\end{figure}

Figure~\ref{fig:acf_temp} shows the ACF (dots) obtained from the EDOS by the Fourier transform of Eq.~\eqref{eq:acf_dos_fourier}.
For all temperatures, one observes a transient decay at short time scales followed by a plateau.
At timescales of order $t \sim 1/\gamma(\beta)$, one observes that the ACF decays further to zero.

Next, we test our simple model for describing the EDOS in terms of the four fitting parameters $(A,\gamma, E_0, w^\star)$, where $A$ and $\gamma$ define the properties of the central Lorentzian and $E_0$ and $w^\star$ define the incoherent sidebands by approximating them with semicircles; see Eq.~\eqref{eq:dos_general_form} and Eq.~\eqref{eq:sidebands_dos_mdl}.

The inset of Fig.~\ref{fig:dos_temp} shows an example of this simple four parameter fit.
In particular, the left inset shows the semi-circle approximation of the incoherent side bands, while the right inset shows a fit to the central Lorentzian peak.

Using the exact ACF of the simplified model
\begin{equation}
    \label{eq:acf_simplified_model}
    \begin{split}
    C_\beta(\hat{O}, t) &= Ae^{-\gamma t}\\ {}&+ (1 - A) \frac{2}{\pi}\frac{J_1(2w^\star t)}{2w^\star t} \cos(E_0t) \, ,
    \end{split}
\end{equation}
where $J_\alpha$ denotes the Bessel function of the first kind, we can compare the predictions from the four parameter fit with the ACF obtained from the Lanczos series.
We find that the simple four parameter fit faithfully catches the short and long timescales of the ACF as can be seen in Fig.~\ref{fig:acf_temp}.

\subsection{\label{sec:results:effective_gaps}Effective Gaps}

\begin{figure}[ht!]
    \centering
    \includegraphics[width=\linewidth]{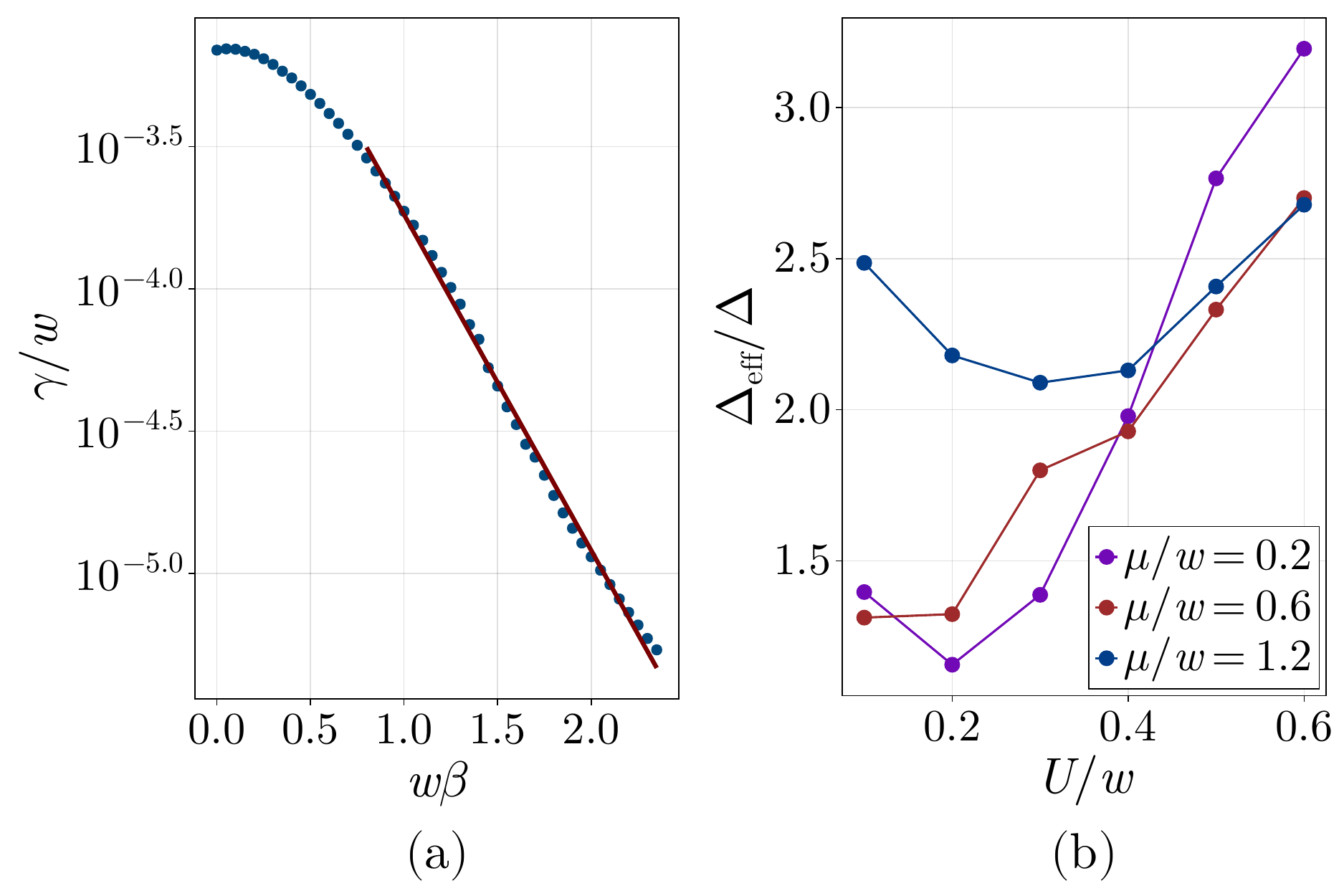}
    \caption{
    (a) An example corresponding to $\mu/w = 1.2$ and $U/w = 0.1$ of the extraction of the effective gap $\Delta_{\rm eff}$ by fitting the lifetime to Eq.~\eqref{eq:effective_gap_definition}.
    (b) Effective gaps along the three cuts $\mu/w = 0.2, 0.6, 1.2$ as shown in Fig.~\ref{fig:phasediagram_sketch}, and normalized by the many-body gaps for each parameter point.}
    \label{fig:fitted_gaps}
\end{figure}

From general arguments, one expects that the lifetime of a local excitation obeys an exponential law~\cite{Sachdev2011}
\begin{equation}
\label{eq:effective_gap_definition}
    \frac{1}{\tau(\beta)} \coloneqq \gamma(\beta) = \gamma_0 e^{-\Delta_{\rm eff}\beta},
\end{equation}
for large $\beta$. 
For a local bulk excitation, $\Delta_{\rm eff}$ is expected to be the many-body gap of the system. Fig~\ref{fig:fitted_gaps}(a) shows the temperature dependence of the inverse lifetime for $\mu/w = 1.2$ and $U/w= 0.1$.

The large $\beta$ regime shows the expected behavior, allowing a fit to be made.
From this fit we obtain an effective energy gap $\Delta_{\rm eff}/w\approx 2.7$, which is significantly larger than the many-body gap of the system ($\Delta/w \approx 1.09$), see Appendix \ref{sec:ap:gap_extraction} for details on how the many-body gap was obtained.
To check that this is not an accidental behavior of the point chosen, we performed the same analysis along the three cuts displayed in Fig.~\ref{fig:phasediagram_sketch}.
The results of the different effective gaps are displayed in Fig.~\ref{fig:fitted_gaps}(b), normalized by the actual many-body gap $\Delta$ of the system.
For every parameter point we observe that the effective gap is larger than $\Delta$ with a non-trivial dependence on the interaction strength $U$. We expect that 
due to the existence of a  SZM that has infinite lifetime at infinite temperature for $U=0$, $\Delta_{\rm eff}/\Delta$ will diverge as $U \rightarrow 0$. The non-monotonic behavior in $U$ where $\Delta_{\rm eff}/\Delta$ also increases
at large $U$ is intriguing and left for future study.

\subsection{Exact Diagonalization}

\begin{figure}[hbt!]
    \centering
    \includegraphics[width=\linewidth]{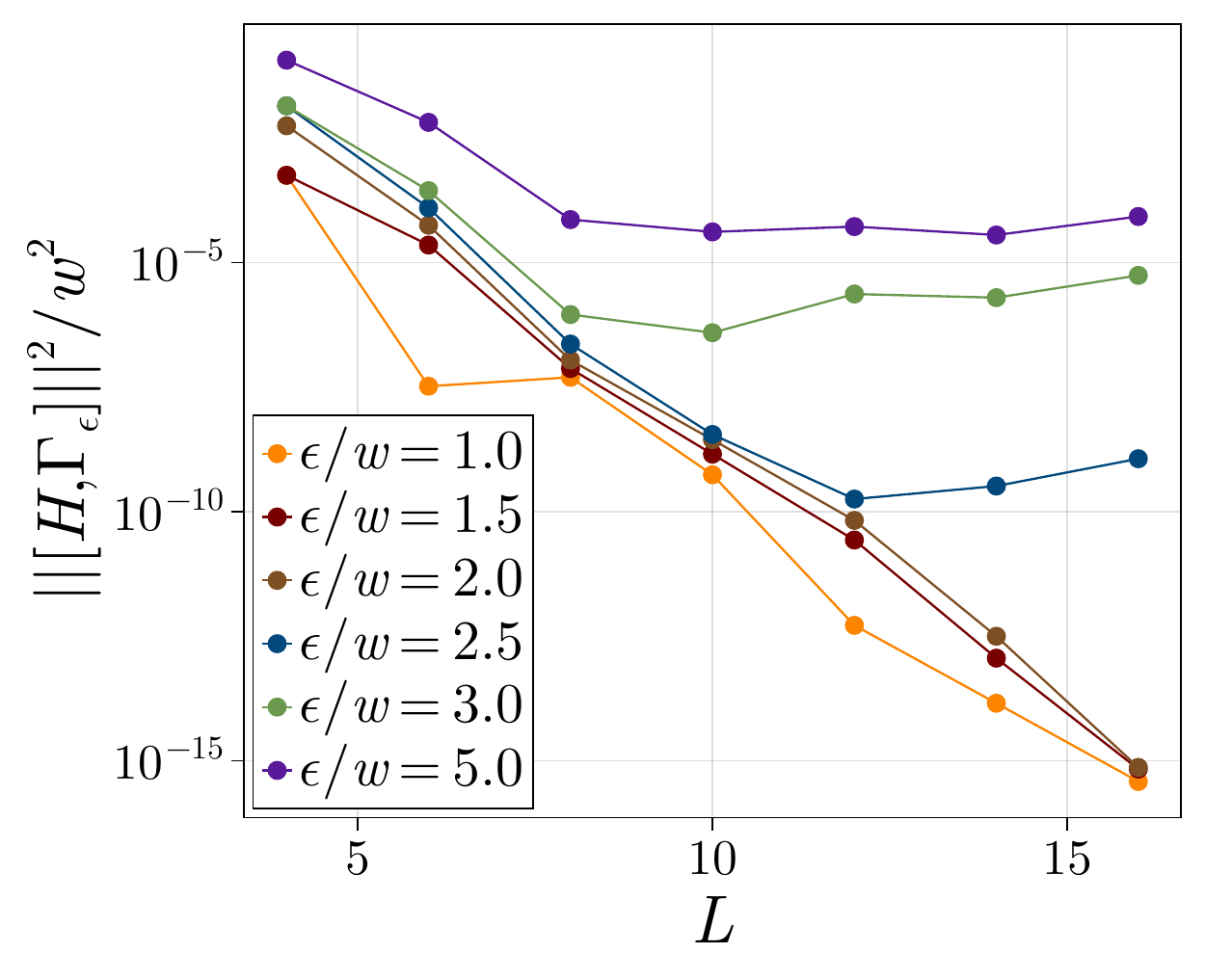}
    \caption{Commutator of the Hamiltonian $H$ with the low energy projected ASZM $\leaszm$, see Eq.~\eqref{eq:szm_cnstr_prj} for $\mu/w = 1.2$ and $U/w = 0.1$.}
    \label{fig:commutator_aszm_ed}
\end{figure}

To check if the effective energy gap obtained in the previous section is reflected in the low energy part of the system, we study the model using exact diagonalization. 
For a given system size $L\le 16$, we calculate the full spectrum $\lbrace\ket{\psi_{n, p}}, E_{n,p}\rbrace$ of the Kitaev-Hubbard chain, with $p=\pm $ being the parity of the state. 
From this, we construct an ASZM as follows~\cite{Kells15}
\begin{equation}
    \label{eq:szm_cnstr}
    \Gamma = \sum_n g_{n}\ket{\psi_{n,+}}\bra{\psi_{n,-}} + \hc\,.
\end{equation}
In the above equation, $g_n\in \U(1)$ is a phase chosen such that $g_n\braket{\psi_{n,-}|\gamma_{1,a}|\psi_{n,+}} \ge 0$, see also Appendix \ref{sec:ap:exact_diagonalization} for a more detailed discussion on the construction.

We can similarly construct an ASZM projected on the low energy sector by
\begin{equation}
    \label{eq:szm_cnstr_prj}
    \begin{split}
    \leaszm &= P_{\epsilon}\, \Gamma\, P_{\epsilon }\\
    {}&=\sideset{}{'}\sum_n g_{n}\ket{\psi_{n,+}}\bra{\psi_{n,-}} + \hc,
    \end{split}
\end{equation}
with the projection operator 
\[
P_{\epsilon}\ket{n,p} = \begin{cases}
    \ket{n,p} \,,\ &\text{for}\ E_{n,p} - E_{0,p} \le \epsilon\\
    0 \,,\ &\text{for}\ E_{n,p} - E_{0,p} > \epsilon
\end{cases} \, .
\]
In Eq.~\eqref{eq:szm_cnstr_prj}, the primed sum means that compared to Eq.~\eqref{eq:szm_cnstr}, we only keep the pairs of states with a maximum excitation energy $\epsilon$ above the ground state.
From $\leaszm$ we can calculate the commutator with the Hamiltonian
\[
||[H,\leaszm]||^2 = \sideset{}{'}\sum_{n} (E_{n,+} - E_{n,-})^2\,.
\]
Another interesting quantity is the overlap of the $\leaszm$ operator with the edge operator $\gamma_{1,a}$.
This overlap is computed with respect to the infinite temperature scalar product, but not normalized by the dimension of the full Hilbert space, but with the dimensionality of the projected space $P_{\epsilon}$.

As an example, we consider $\mu/w = 1.2$, $U/w = 0.1$.
In the previous section, from the temperature dependence of the Lanczos coefficients, we obtained an effective energy gap of $\Delta_{\rm eff}/w \approx  2.7$.
In Fig.~\ref{fig:commutator_aszm_ed} we plot the commutator of the low energy projected ASZM for even system sizes $4 \le L \le 16$. 
We find that for $\epsilon/w< 2.5$ the commutator shows an exponentially decaying behavior with $L$,  with approximately the same slope.
This behavior changes qualitatively for $\epsilon/w > 2.5$, where the slope is much smaller, with the commutator reaching a $L$ independent plateau value. 
The behavior changes around $\epsilon/w \sim 2.5$ close to the effective energy gap $\Delta_{\rm eff}/w \approx  2.7$ obtained from the temperature analysis of the Lanczos coefficients.

\begin{figure}[hbt!]
    \centering
    \includegraphics[width=\linewidth]{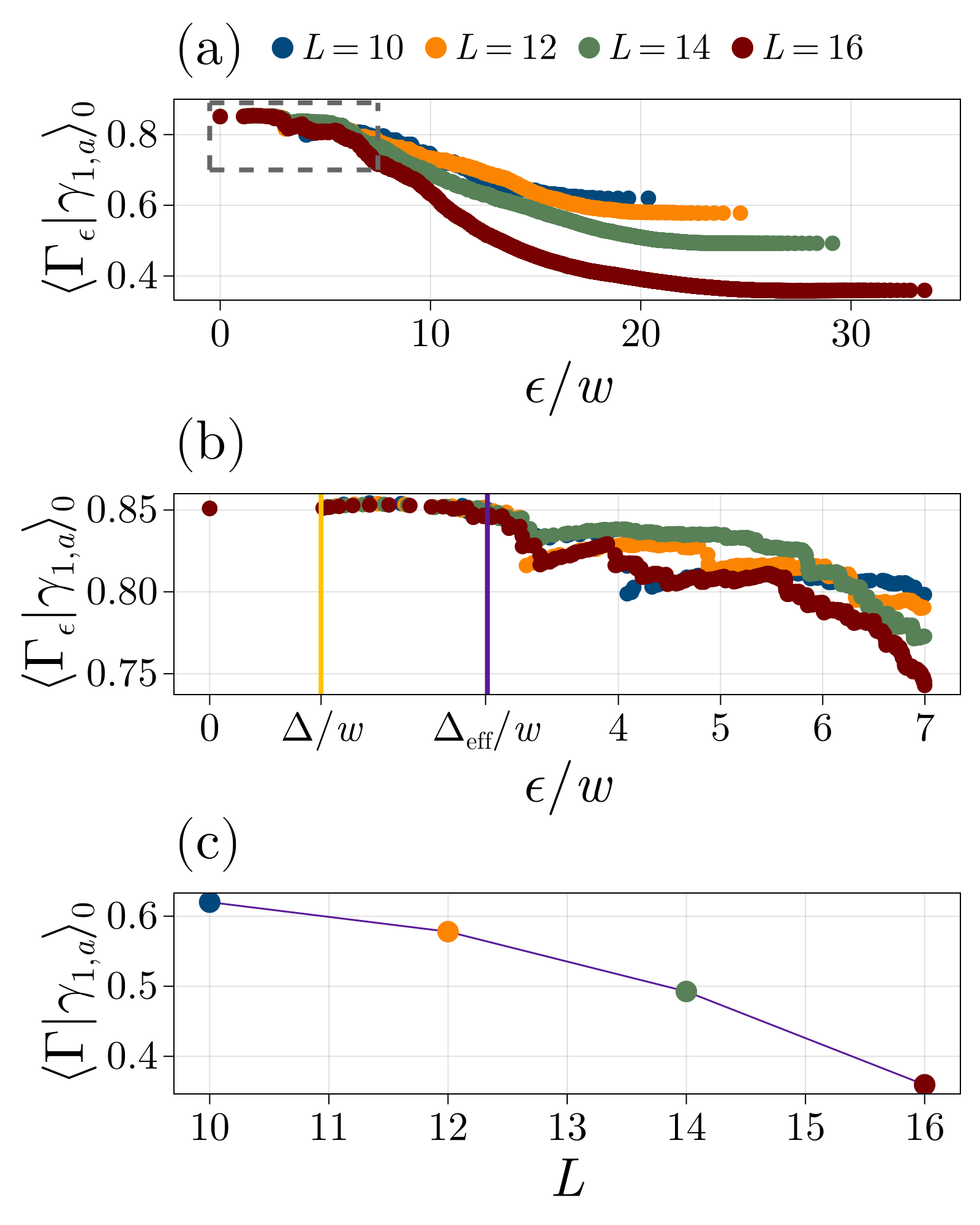}
    \caption{
    Overlap of the projected ASZM $\leaszm$ with the edge operator $\gamma_{1,a}$ for $\mu/w = 1.2$ and $U/w = 0.1$. (a) Varying the energy cutoff $\epsilon$ over all scales. For large enough $\epsilon$, we observe a decay of the overlap with respect to the system size. (b) A detailed plot for all energies $\epsilon/w < 7$, corresponding to the gray dotted box shown in a). The effective energy gap $\Delta_{\rm eff}$ is highlighted by a purple vertical line, while
    the thermodynamic many-body gap $\Delta$ is denoted by the yellow vertical line. (c) Plot of the overlap with the full ASZM $\Gamma$ for different system sizes. All three plots have the same color code for the system size $L$.}
    \label{fig:overlaps_example}
\end{figure}

Next, we consider the overlap of $\leaszm$ with the edge operator $\gamma_{1,a}$.
This overlap is plotted against the cutoff energy $\epsilon$ in Fig.~\ref{fig:overlaps_example}.
If we do not impose any cutoff, i.e., considering the full ASZM $\Gamma$, we observe that the overlap decays with the system size.
This is explicitly shown in Fig.~\ref{fig:overlaps_example}(c).
However, by reducing the cutoff energy, we observe that for $\epsilon \lesssim \Delta_{\rm eff}$ the value of the overlap appears to converge with the system size.
Combining the results for the commutator $||[H,\leaszm]||^2$ with the results for the overlap
$\langle\leaszm|\gamma_{1,a} \rangle_0$
, we conclude that if we project the system to an energy below $\Delta_{\rm eff}$, we observe the emergence of an operator $\leaszm$ for which the commutator with the Hamiltonian $H$ vanishes exponentially in the system size, while having a finite overlap with the edge operator $\gamma_{1,a}$.
In that sense, $\leaszm$ becomes a strong zero mode for the low energy sector of the Kitaev-Hubbard chain.

\section{\label{sec:conlusion}Conclusions}

An important topic both from a theoretical perspective as well as for practical realizations of quantum memories, is understanding the stability of topologically protected edge modes when interactions are present, and the system does not lie in the ground state sector.
Our work takes a step in this direction by interpolating between zero temperature and infinite temperature. 
We find that, quite remarkably, the topological protection in the ground state manifold may not vanish immediately on raising the temperature, with stable edge modes present in an energy window which is of the order of, but systematically larger than the many-body gap. 
In arriving at this result, we have combined two methods for studying operator dynamics, Lanczos series expansions and tensor network ansatz, thus allowing us to access dynamics in the notoriously difficult regime of excited states, long times, and large systems.  

While the particular example studied in this work was an interacting fermionic chain protected only by the $\Z_2$ fermion parity, the approach can be easily extended to various other systems realizing stable edge modes.
These include parafermionic systems~\cite{Fendley2012, Alicea2016, Iemini2017_B} protected by general $\Z_n$ symmetries and Floquet circuits, which host Majorana modes and the more exotic $\pi$ modes~\cite{Sen13, Yates2019,Harper2020, Yates2021,Matthies2022}.
Another interesting direction is the study of Majorana edge modes realized by quasi-one dimensional fermionic ladder systems with strong pair hopping between the two ladders~\cite{Cheng2011, Kraus2013, Lang2015, Iemini2017_B}.
The topological phase in this ladder system occurs without breaking the total particle number conservation, opening the possibility for experimental realizations~\cite{Iemini2017, Lisandrini2022, Tausendpfund2023, Defossez2024, Michen2024}.
This additional global $U(1)$ symmetry comes at the cost of gapless density fluctuations~\cite{Cheng2011,Sau2011,Fidkowski2011} which makes the stability of these edge modes at finite temperatures questionable.

Other future directions include understanding the precise transition from absolutely stable SZMs at low energies to unstable but long-lived ASZMs at high energies.
In this regard, it may be interesting to study the interplay of disorder and interactions, as it is possible that disorder increases the region of the spectrum that hosts SZMs~\cite{Kells2018}, which is related to the phenomenology of many body localization.
Finally, a fruitful direction of research is a more efficient construction of ASZMs by employing variational approaches tailored to directly target the low energy space of a theory.

\section{Acknowledgments}

We acknowledge fruitful discussions with D. Alcalde, S. Diehl, R. Egger, G. Kells, A. Rosch, S. Trebst, E. Weerda.
The simulations presented in this work were produced with a code based on the \rm{ITensor} library~\cite{Fishmann2022}. Data and code are available at~\cite{Tausendpfund2025_dataset}.
N.T. and M.R. acknowledge the support from the DFG under Germany's Excellence Strategy - Cluster of Excellence Matter and Light for Quantum Computing (ML4Q) EXC 2004\slash 1 – 390534769 and project Grant No. 277101999 within the CRC network TR 183. 
A.M. acknowledges the support of the US Department of Energy, Office of Science, Basic Energy Sciences, under Award No.~DE-SC0010821.
The authors gratefully acknowledge the Gauss Centre for Supercomputing e.V. (www.gauss-centre.eu) for funding this project by providing computing time through the John von Neumann Institute for Computing (NIC) on the GCS Supercomputer JUWELS~\cite{JUWELS} and through FZJ on JURECA~\cite{JURECA2021} at J\"ulich Supercomputing Centre (JSC).

\appendix

\section{\label{sec:ap:ftsclprd}Finite Temperature Scalar Product}

In this paper, we consider a many-body Hilbert space $\hilb$ defined on a finite chain.
On this Hilbert space we consider the set of all operators $\hat{O}:\hilb \to \hilb$.
Due to linearity, this set is itself a vector space denoted by $\hilb^{\rm op}$.
It is also possible to define a scalar product on $\hilb^{\rm op}$.
A general class of possible choices for a valid scalar product defined for a finite temperature $T = 1/\beta$ is given by~\cite{viswanath1994recursion}
\begin{equation}
    \label{eq:basic_notions:scalar_prd_t_gen}
    \braket{A|B}_\beta^g \coloneqq \frac{1}{\partf} \int_0^\beta \!\der\lambda\, g(\lambda) 
    \Tr\left[
        y^{\beta - \lambda} A^\dag y^{\lambda} B
    \right]\, .
\end{equation}
Here 
\[
    y \coloneqq e^{-H}\, , \quad \partf = \Tr\left[ e^{-\beta H} \right],
\]
and $g: [0, \beta] \to \mathbb{R}_+$ is a positive function with the properties
\[
    \frac{1}{\beta}\int_0^\beta \!\der \lambda g(\lambda) = 1\,,\quad g(\beta - \lambda) = g(\lambda)\,.
\]

There are two important choices for the function $g(\lambda)$
\[
    \begin{split}
        g_S(\lambda) &= \frac{1}{2} \left(
            \delta(\lambda) + \lambda(\beta - \lambda)
        \right)\, ,\\\ g_W(\lambda) &= \delta(\beta/2 - \lambda),
    \end{split}
\]
leading to the two finite temperature scalar products:
\begin{equation}
    \begin{split}
    \label{eq:basic_notions:autocorr_finite_temp}
    \braket{A|B}_\beta^S &= \frac{1}{2}\Tr\left[
        \rho(\beta)\lbrace A^\dag B + B A^\dag \rbrace
    \right]\,,\\
    \braket{A|B}_\beta^W &= \frac{1}{\partf} \Tr\left[
        e^{-\frac{\beta}{2} H} A^\dag e^{-\frac{\beta}{2} H} B
    \right] \, .
    \end{split}
\end{equation}
Above $\rho(\beta) = e^{-\beta H}/Z(\beta)$.
The first choice naturally appears in linear response theory while the second choice is related to Wightman correlation functions~\cite{Parker2019, viswanath1994recursion}.

\section{\label{sec:ap:JWT}Jordan-Wigner Transformation}

The Jordan-Wigner transformation~\cite{Jordan1928, Lieb1961} is a non-local unitary transformation of the Hilbert-space that maps fermionic degrees of freedom to spins.
Let $\mathcal{H}$ again denote a many-body fermionic Hilbert-space generated from the vacuum $\ket{0}$ by the set of fermionic operators $\lbrace c_j^\vdag \rbrace$ obeying the canonical anticommutation relations.

The local Hilbert-space $\mathcal{H}^{\rm loc}_j$ is formed by the two states $\ket{0}_j$, which is the vacuum, and $\ket{1}_j = c_j^\dag\ket{0}$, which hosts one fermionic particle.
The Jordan-Wigner transformation acts on this local Hilbert-space by identifying the states
\[
\ket{0}_j \to \ket{\downarrow}_j\,,\quad \ket{1}_j \to \ket{\uparrow}_j,
\]
together with the transformation of operators:
\begin{equation*}
    \begin{split}
        c_j^\vdag &= \frac{1}{2}\left(\sigma_j^x - i\sigma_j^y \right)  \mathcal{S}_j\,, \
        c_j^\dag  = \frac{1}{2}\left(\sigma_j^x + i\sigma_j^y \right)  \mathcal{S}_j, \, \\
        n^\vdag_j &= \frac{1}{2}\left( \sigma_j^z + 1\right)\,,\, 
        \mathcal{S}_j
        = \prod_{k<j}\left(-\sigma_k^z\right).
    \end{split}
\end{equation*}
Consider now the Kitaev-Hubbard chain with a general p-wave pairing potential
\begin{equation}
    \label{eq:kitaev_hubbard_chain_extended}
    \begin{split}
         H = \sum_{j = 1}^{L-1} &\, -w c_j^\dag c_{j+1}^\vdag -\Delta c_j^\dag c_{j+1}^\dag +\hc \\
   {} &+ U\, \sum_{j = 1}^{L-1} p_j p_{j+1} - \frac{\mu}{2}\,\sum_{j = 1}^L p_j \, ,
    \end{split}
\end{equation}
which reduces to Eq.~\eqref{eq:kitaev_hubbard_chain} considered in the main text for $\Delta = w$.

Applying the Jordan-Wigner transformation to this Hamiltonian leads to the XYZ spin chain in a magnetic field
\begin{equation}
    \label{eq:xyz_ham}
    \begin{split}
    H =& \sum_{j = 1}^{L-1} \left[
        J_x \paulx{j}\paulx{j+1} +
        J_y \pauly{j}\pauly{j+1} +
        J_z \paulz{j}\paulz{j+1}
    \right]\\
    {}&+ g\,\sum_{j = 1}^{L} \paulz{j}\,.
    \end{split}
\end{equation}
The parameters are identified using
\begin{equation*}
    \begin{split}
    J_x &= -\frac{w + \Delta}{2}\,,\ J_y = -\frac{w - \Delta}{2}\,,\\
    J_z &= U\,,\ g = -\mu/2 \, .
    \end{split}
\end{equation*}
Under the Jordan-Wigner transformation, the fermionic parity $P = \exp(i\pi \sum_{j=1}^L n_j)$ becomes the product over all $\sigma_j^z$.

\section{\label{sec:ap:jw_sec_line}Solution of the Kitaev-Hubbard chain at \texorpdfstring{$\mu=0$}{mu=0}}

In this appendix, we discuss the transformation to diagonalize the Kitaev-Hubbard chain for $\mu = 0$.
The Hamiltonian of Eq.~\eqref{eq:kitaev_hubbard_chain} reduces to
\begin{equation}
    \begin{split}
    H = \sum_{j = 1}^{L-1} &-w\left(c_j^\dag - c_j^\vdag \right)
         \left(c_{j+1}^\dag + c_{j+1}^\vdag\right) \\
        + &U p_j p_{j+1} \, .
    \end{split}
\end{equation}
Using the Jordan-Wigner transformation as described in Appendix \ref{sec:ap:JWT}, this Hamiltonian becomes
\begin{equation}\label{eq:jw_trs_ham_1}
    H = \sum_{j = 1}^{L-1} \left[
        -w \paulx{j}\paulx{j+1} +
         U \paulz{j}\paulz{j+1}
    \right]\,.
\end{equation}
Now performing a second Jordan-Wigner transformation switches the role of $\sigma_j^y$ with $\sigma_j^z$
\begin{equation*}
    \begin{split}
        f_j^\vdag &= \frac{1}{2}\left(\sigma_j^x - i\sigma_j^z \right)  \mathcal{S}^\prime_j\,, \ 
        \mathcal{S}_j^\prime
        = \prod_{k<j}\left(-\sigma_k^y\right) \, .
    \end{split}
\end{equation*}
With this transformation, the Hamiltonian~\eqref{eq:jw_trs_ham_1} becomes
\[
H = \sum_{j=1}^{L-1} -\tilde{w} f_j^\dag f_{j+1}^\vdag -\tilde{\Delta} f_{j}^\dag f_{j+1}^\dag + \hc\, ,
\]
which is again of type~\eqref{eq:kitaev_hubbard_chain_extended} with zero chemical potential and interaction. 
One has $\tilde{w} = w - U$ and $\tilde{\Delta} = w + U$. 
This model is known to have a SZM, which for $L\to \infty$ is
\begin{equation}
\begin{split}
    \szm &= \mathcal{N}\sum_{k = 1}^{\lfloor \frac{L+1}{2}\rfloor} \left(\frac{U}{w}\right)^{k-1}\!\! (f_{2k-1}^\dag + f_{2k-1}^\vdag)\,,\\
    \mathcal{N} &= {\sqrt{1 - (U/w)^2}}\,.
    \end{split}
\end{equation}

Let us now rewrite this in terms of the original fermions $c_j$.
First note that the Jordan-Wigner strings $\mathcal{S}_{2k-1}^\prime$ are given in terms of the Majorana operators $c_j = (\gamma_j^a -i\gamma_j^b)/2$ as
\[
\mathcal{S}_{2k-1}^\prime = 
i^{k-1} \gamma_1^a \gamma_2^b \gamma_3^a \dots \gamma_{2k-2}^b \,,
\]
which can be proven by induction. In the original fermions, the SZM is then given by 
\begin{equation}
    \szm = \mathcal{N} \sum_{k=1}^{\lfloor \frac{L+1}{2}\rfloor}\left(\frac{U}{w}\right)^{k-1} \!\!
    \mathcal{P}_k \gamma_{2k-1}^a \, ,
\end{equation}
where $\mathcal{P}_k$ is the string operator
\[
\mathcal{P}_k = \mathcal{S}_{2k-1}^\prime \mathcal{S}_{2k-1} = i^{k-1} \gamma_1^b \gamma_2^a \gamma_3^b \dots \gamma_{2k-2}^a \,.
\]

\section{\label{sec:ap:acf_greensfunctions_cntfrct}Autocorrelation function from the Greens function}

In this appendix, we give more details on how the EDOS can be calculated efficiently by the continued fraction technique.
The EDOS is defined as
\[
    \nu_\beta^E(\omega) = -\frac{1}{\pi}\lim_{\eta \to 0^+}\mathcal{I}\left[
            G_\beta^E(\omega + i\eta)
        \right],
\]
where $\mathcal{I}[z]$ is the imaginary part of a complex number $z$ and 
\begin{equation}\label{eq:edge_greensfunction}
    G_\beta^E(z) = -i\braket{1|\frac{1}{z\unit - H_{\rm sp}}|1},
\end{equation}
is the edge Greens function of the artificial single particle Hamiltonian $H_{\rm sp}$. 
This expression for the EDOS is equivalent to the expression given in the main text of equation~\eqref{eq:edge_dos}.

We assume the following structure of $H_{\rm sp}$
\[
    H_{sp} = H_0 + H_1 + V + V^\dagger,
\]
with
\[
\begin{split}
    H_0 &= \sum_{n=1}^{N-1} -t_n \ket{n+1}\bra{n} +\hc\,,\\ H_1 &= \sum_{n = N+1}^\infty -t_n \ket{n+1}\bra{n} +\hc,
\end{split}
\]
and $V = -t_\star \ket{N}\bra{N+1}$.
The part given by $H_0$ contains all the hopping amplitudes obtained by the Lanczos series $b_n = it_n$ as outlined in section~\ref{sec:lanzcos_series}.
The second part $H_1$ represents all the unknown hopping amplitudes which we interpolate by choosing a suitable model, and $V$ represents the coupling between the two parts.
As the Lanczos coefficients typically reach a plateau value for some $n_\star$, we choose $t_n = t_\star = -ib_{N}$ for $n>N > n_\star$.
Thus, we model the unknown Lanczos coefficients by a semi-infinite homogeneous chain.

To evaluate the edge Greens function~\eqref{eq:edge_greensfunction} we make use of the block inversion formula \cite{Abadir2005}
\begin{equation}\label{eq:blck_inv_frm}
P_A\begin{pmatrix}
    H_1 & V \\
    W & H_2
\end{pmatrix}^{-1}P_A = (H_1 - VH_2^{-1}W)^{-1},
\end{equation}
with $P_A$ being a projector on the first diagonal block.
Applied to the edge Greens function one finds
\begin{equation}\label{eq:edge_greensfunction_2}
    iG_\beta^E(z) = \left(
        z\unit - H_0 - |t_\star|^2 \tilde{G}_\beta^E(z) \ket{N}\bra{N}
    \right)^{-1}_{1,1},
\end{equation}
where $\tilde{G}_\beta^E(z)$ is the EDOS of $H_1$ and can be calculated analytically for the homogeneous chain
\begin{equation}
    \tilde{G}_\beta^E(z) = \frac{1}{4|t_\star|^2}\left(
z + \sqrt{z^2 - 4|t_\star|^2}
\right)\,.
\end{equation}

The equation~\eqref{eq:edge_greensfunction_2} can now be evaluated by explicit inversion.
Alternatively, one can again use the block inversion formula~\eqref{eq:blck_inv_frm} to further reduce the expression, obtaining the finite continued fraction
\begin{equation}
    iG_\beta^E(z) = \cfrac{1}{
    z - \cfrac{|t_1|^2}{
        z - \cfrac{|t_2|^2}{
                z - \cfrac{\dots}{z - |t_\star|^2\tilde{G}^E_\beta(z)}
            }
        }
    }\, ,
\end{equation}
which is numerically more stable and faster to compute than the explicit numerical inversion of $z\unit - H_0 - |t_\star|^2 \tilde{G}_\beta^E(z)$.

\section{\label{sec:app:edos_ssh_connection}Approximate Edge Density of States (EDOS)}

In this appendix we review the continuum approximation of the artificial Hamiltonian.
The discussion mainly follows~\cite{Yates2020_B}.
We will employ the continuum description reviewed here to motivate the phenomenological fit of Eq.~\eqref{eq:sidebands_dos_mdl}.

We start from the artificial single particle Hamiltonian
\[
    H = \sum_{n=0}^N ib_{n+1}\ket{n+1}\bra{n} + \hc,
\]
which gives the Schr\"{o}dinger time evolution
\begin{equation}\label{eq:schrd_eq_app}
    i\partial_t \varphi_n(t) = ib_n \varphi_{n-1}(t) - ib_{n+1}\varphi_{n+1}(t),
\end{equation}
with $\varphi_n(0) = \delta_{n,1}$. We assume the form $b_n =  h_n + (-1)^n \tilde{h}_n$, where the slowly varying $k=0,\pi$ components are $h_n$ and $\tilde{h}_n$, respectively.
This allows for a splitting of the wavefunction $\varphi_n = \varphi_n^0 + (-1)^n \varphi_n^\pi$.
The $k=0,\pi$ components of the wavefunction can now be approximated by the envelope function
\[
    \varphi^k_n \approx \varphi^k(an)\, ,
\]
which is assumed to be smooth and to vary slowly on the length scale of $a$, representing the lattice spacing between sites $n$ and $n+1$.
Introducing the Dirac spinor $\Psi(x) = (\varphi^0(x),\varphi^\pi(x))^T$, the Schr\"{o}dinger equation~\eqref{eq:schrd_eq_app} can be approximated by the Dirac-like equation
\begin{equation}
    i\partial_t \Psi = -a \lbrace h(x), i\partial_x \rbrace \sigma_z \Psi + m(x) \sigma_y \Psi,   
\end{equation}
with the mass term $m(x) = 2\tilde{h}(x) + a\partial_x \tilde{h}(x)$, and $h(x)$ and $\tilde{h}(x)$ are smooth approximations to $h_n$ and $\tilde{h}_n$.

To remove the position dependence of the momentum operator, we consider the general coordinate transformation $y = y(x)$, with
\begin{equation}\label{eq:ap:coordinate_traffo_1}
   \partial_x y = \frac{1}{2 h(x)}\,,\ y(a) = a\,.
\end{equation}
The initial condition is chosen so that the chain starts at the same point in both the original and transformed coordinates.
This transformation is bijective as long as $h(x)>0$. Defining the rescaled Dirac spinor $\chi = \sqrt{h} \Psi$, the transformed Dirac equation reads
\begin{equation}
\label{ap:eq:chi_1}
    i\partial_t \chi = \left[-a i \partial_y \sigma_z + \tilde{m}(y) \sigma_y \right]\chi\,.
\end{equation}
This equation now resembles a standard Dirac equation with the position dependent mass term $\tilde{m}(y) = 2\tilde{h}(y) + \partial_y \tilde{h}(y)/(2h(y))$.

We can now try to find a second lattice Hamiltonian $\tilde{H}_{\rm sp}$ which has the same continuum limit as given by equation~\eqref{ap:eq:chi_1}. 
For this consider
\begin{equation}
    \tilde{H}_{\rm sp} = \sum_{y=0}^\infty -it_{n+1}\ket{n+1}\bra{n} +\hc,
\end{equation}
where $t_{n} = t + (-1)^n \tilde{t}_n$.
Note that $t$ is now a constant. The new staggered component $\tilde{t}_n$ is again assumed to vary slowly.
The continuum version of the Schr\"{o}dinger equation given by $\tilde{H}_{\rm sp}$ is now
\begin{equation}
\label{ap:eq:chi_2}
    i\partial_t \chi = \left[-i 2t a \partial_y \sigma_z + M(y) \sigma_y \right]\chi\, ,
\end{equation}
with $M(y) =  2\tilde{t}(y) + \partial_y \tilde{t}(y)$.
Since $t$ was a constant to start with, the momentum part has no additional position dependence.

Comparing equations~\eqref{ap:eq:chi_1} and \eqref{ap:eq:chi_2}, we find both lead to the same time evolution if we set $t = 1/2$ and
\begin{equation}\label{eq:ap:differential_transformed_model}
    2\tilde{t}(y) + \partial_y \tilde{t}(y) =
    2\tilde{h}(y) + \partial_y \frac{\tilde{h}(y)}{2h(y)}\, .
\end{equation}
We choose the initial condition to be $\tilde{t}(a) = \tilde{h}_1$.
In general, this first-order differential equation has a unique solution that completely fixes $\tilde{t}(x)$ by the parameters of the original model, $h$ and $\tilde{h}$.
An approximate solution can be found by discarding the derivatives on both sites of Eq.~\eqref{eq:ap:differential_transformed_model}
\[
    \tilde{t}(y) = \tilde{h}(x(y)),
\]
where $x(y)$ is the inverse of the transformation defined in Eq.~\eqref{eq:ap:coordinate_traffo_1}.
We can now set $v(y) = 1/2 - \tilde{h}(x(y))$ and $w(y) = 1/2 + \tilde{h}(x(y))$ such that the new lattice Hamiltonian $\tilde{H}_{\rm sp}$ reads
\[
\begin{split}
\tilde{H}_{\rm sp} = \sum_n &v(an) \ket{n, a} \bra{n, b}\\
    {}&+ w(an) \ket{n,b}\bra{n+1,a} + \hc\,.
\end{split}
\]
The single particle Hamiltonian $\tilde{H}_{\rm sp}$ is thus given by a SSH chain with position dependent staggering of the hopping amplitudes centered around the constant value of $1/2$.
Within this approximation, knowing the solution of the Schr\"{o}dinger equation for $\chi$ derived from $\tilde{H}_{\rm sp}$ is equivalent to knowing the solution of the original Schr{\"o}dinger equation in terms of the variables $\varphi_n(t)$, and derived by $H_{\rm sp}$.
Explicitly focusing on the first site, one has
\[
C(t) = \braket{1|e^{-itH_{\rm sp}}|1} \approx \braket{1|e^{-it\tilde{H}_{\rm sp}}|1} \, .
\]
Furthermore, we find that the EDOS of $H_{\rm sp}$ is the same as the EDOS of $\tilde{H}_{\rm sp}$.

We now consider a simple model with a constant staggering up to $n_\star$, and with the background $h$ increasing linearly with $n$:
\[
h(n) = \alpha + bn\,,\ \tilde{h}_n = \rho \theta(n_\star - n)\,.
\]
Here one finds for the transformed variables:
\[
    \begin{split}
        y &= \frac{1}{b} \log\left(\frac{\alpha + bn}{\alpha + b}\right) + 1,\\
        x &= \frac{1}{b} \left( 
            e^{b(y-1)}(\alpha + b) - \alpha
        \right).        
    \end{split}
\]
The transformed model describes a short SSH chain in the topological regime $v = 1/2-\rho< w = 1/2 + \rho$ until it reaches the location $y_{\star} = y(n_{\star})$, after which it becomes metallic.
In such a situation, the topological edge mode in the SSH chain overlaps with the metallic bulk, leading to edge mode leakage.
This leakage leads to a broadening of the delta peak at zero energy in the EDOS of a pure SSH chain.
In contrast, the side bands of the EDOS describe the hybridization of the bulk bands of the SSH chain with the metallic states of the lead, and results in only small modifications of the density of states.

\begin{figure}[htb!]
    \centering
    \includegraphics[width=\linewidth]{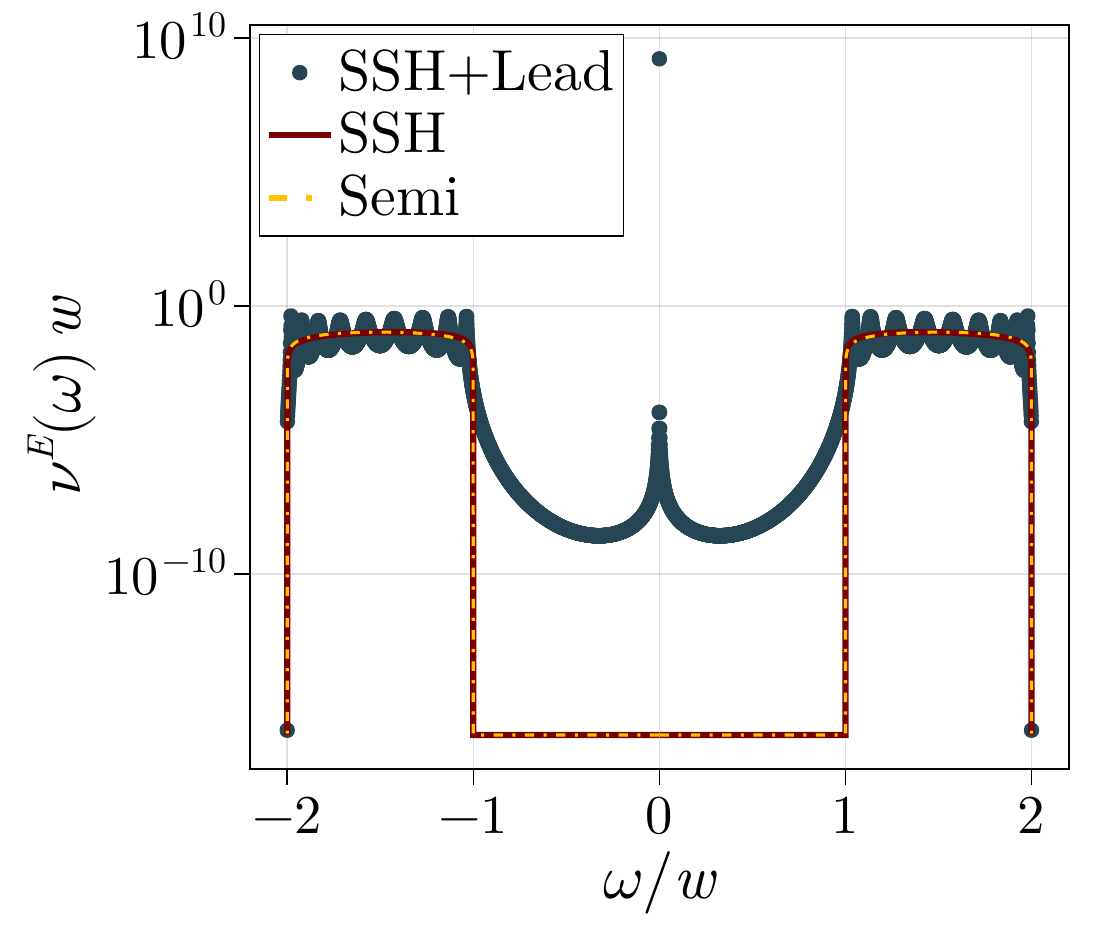}
    \caption{EDOS obtained for a short SSH chain with $v = 0.5$, $w = 1.5$ of length $L = 20$ attached to a semi-infinite homogeneous chain with hopping $t = 1$. The red line is the bulk contribution of the EDOS of a semi-infinite SSH chain with the same parameters, the orange dashed line is the semi-circle approximation of the side bands and it  completely overlaps the red line of the SSH EDOS.}
    \label{fig:short_ssh_chain}
\end{figure}

The exact EDOS of the SSH chain with hopping parameters $v>0$ and $w>0$ is given by
\begin{equation}\label{eq:edos_ssh}
\begin{split}
    \nu_E(\omega) &= \frac{w^2 - v^2}{w^2}\delta(\omega)\Theta(w - v)\\
    +&\frac{1}{2\pi \omega w^2} \sqrt{4v^2w^2 - (\omega^2 -v^2 -w^2)^2}\\
    \times&\Theta(|\omega| - |v-w|) \Theta(|v+w| - |\omega|).
\end{split}
\end{equation}
The first line of Eq.~\eqref{eq:edos_ssh} is due to the topological edge state present for $v<w$, while the second line describes the bulk contribution to the EDOS.
The bulk gap of this SSH chain is given by $\Delta = |v - w|$, the bands are centered around $\pm\bar{E}_0 = \pm \max(v,w)$, and the bandwidth is given by $\delta = \min(v,w)$.
In the large gap limit, the side bands are well approximated by simple semi-circles
\begin{equation}\label{eq:edos_semicircle_app}
    \nu_E(\omega) = \frac{A}{\pi \delta}\sqrt{1 - \frac{(|\omega| - \bar{E}_0)^2}{\delta^2}},
\end{equation}
for $|v - w| \le |\omega| \le |v + w|$. 
The factor $A$ is associated with the spectral weight of the potential edge mode.
One has $A = (v/w)^2$ for $v<w$ (topological) and $A = 1$ for $v>w$ (trivial).
To demonstrate this behavior, we plot the EDOS of a short SSH chain attached to a homogeneous lead in Fig.~\ref{fig:short_ssh_chain}, and compare it to the EDOS of a semi-infinite SSH chain, Eq.~\eqref{eq:edos_ssh}, and  to the semi-circle approximation of Eq.~\eqref{eq:edos_semicircle_app}.

\section{\label{sec:ap:more_numerical_results}Further numerical results}

In this appendix we present additional numerical results.

\subsection{\label{sec:ap:convergence}Convergence Properties}

To calculate the EDOS and thus the lifetime of the edge modes at finite temperatures, we made two approximations: 
The truncation of the bond dimension of the matrix product operators, and the modeling of the unknown Lanczos coefficients by a semi-infinite homogeneous chain.
In the following we discuss how these two approximations influence the results.

In section~\ref{sec:tn_lanczos} we outlined that we approximate the orthonormal basis operators $\mathcal{O}_n$ obtained from the Lanczos series by a matrix product operator with a finite bond dimension $\chi$.
\begin{figure}[htb!]
    \centering
    \includegraphics[width=\linewidth]{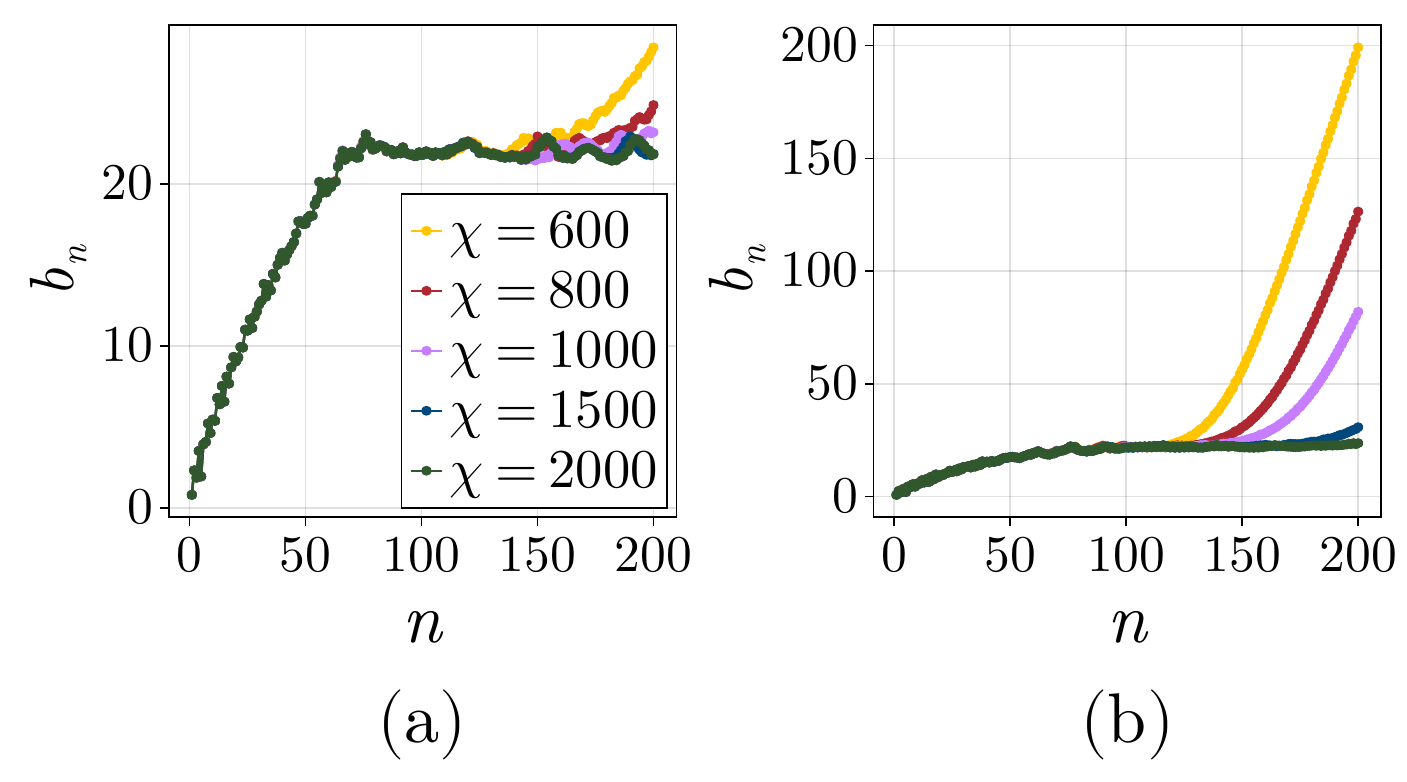}
    \caption{Lanczos coefficients $b_n$  obtained for $\mu/w = 0.6,\ U/w = 0.3$, and for a system of size $L=22$, for different bond dimensions. Panel (a) [(b)] is for an  inverse temperature $w\beta = 0.4$ ($w\beta = 1.25$).}
    \label{fig:convergence_lanczos}
\end{figure}
A finite bond dimension limits the amount of independent operators that can be present in $\mathcal{O}_n$.
As a result, the sequence of Lanczos coefficients $b_n$ is not precise, but only an approximation for any given value of $\chi$.
Fig.~\ref{fig:convergence_lanczos} shows the Lanczos coefficients for two different inverse temperatures $t\beta = 0.4$ (left panel) and $w\beta = 1.25$ (right panel), and for different bond dimensions, for the parameters $\mu/w = 0.6, U/w = 0.3$. 
At both temperatures we observe that the Lanczos series becomes unstable for some value of $n_{\rm unstable}$.
For larger $\beta$ (smaller $T$) the instabilty occurs for a smaller value of $n_{\rm unstable}$.
This can be understood by noting that a large part of the Hilbert space has an exponentially small weight at sufficiently small temperatures.
Small numerical errors can quickly accumulate and lead to instability of the Lanczos iteration.
The situation is worst at strictly zero temperature, where the scalar product of Eq.~\eqref{eq:ft_sclprd} actually has a large null space.
It is thus a pseudo scalar product instead of a real scalar product. 
At infinite precision, this null space should be projected out of the Lanczos series.
However, small numerical errors can lead to large contributions within this null space.
Increasing the bond dimension $\chi$ reduces  numerical errors, thereby shifting this instability to larger values of $n$.
This is a generic feature obtained for every parameter combination $(\mu/w, U/w)$ we have studied in this work. 

To obtain any meaningful results from the $b_n$, it is therefore crucial to truncate the Lanczos series before the instability sets in.
For example, for the $\chi=400$ and $w\beta = 0.4$, one truncates the $b_n$ at $n \sim 100$. 
However, it is crucial that the instability occurs at a value $n_{\rm unstable}$ that is deep inside the plateau.
In addition, in order to obtain a faithful result for the lifetime, the staggered component should already have decayed.
\begin{figure}[htb!]
    \centering
    \includegraphics[width=\linewidth]{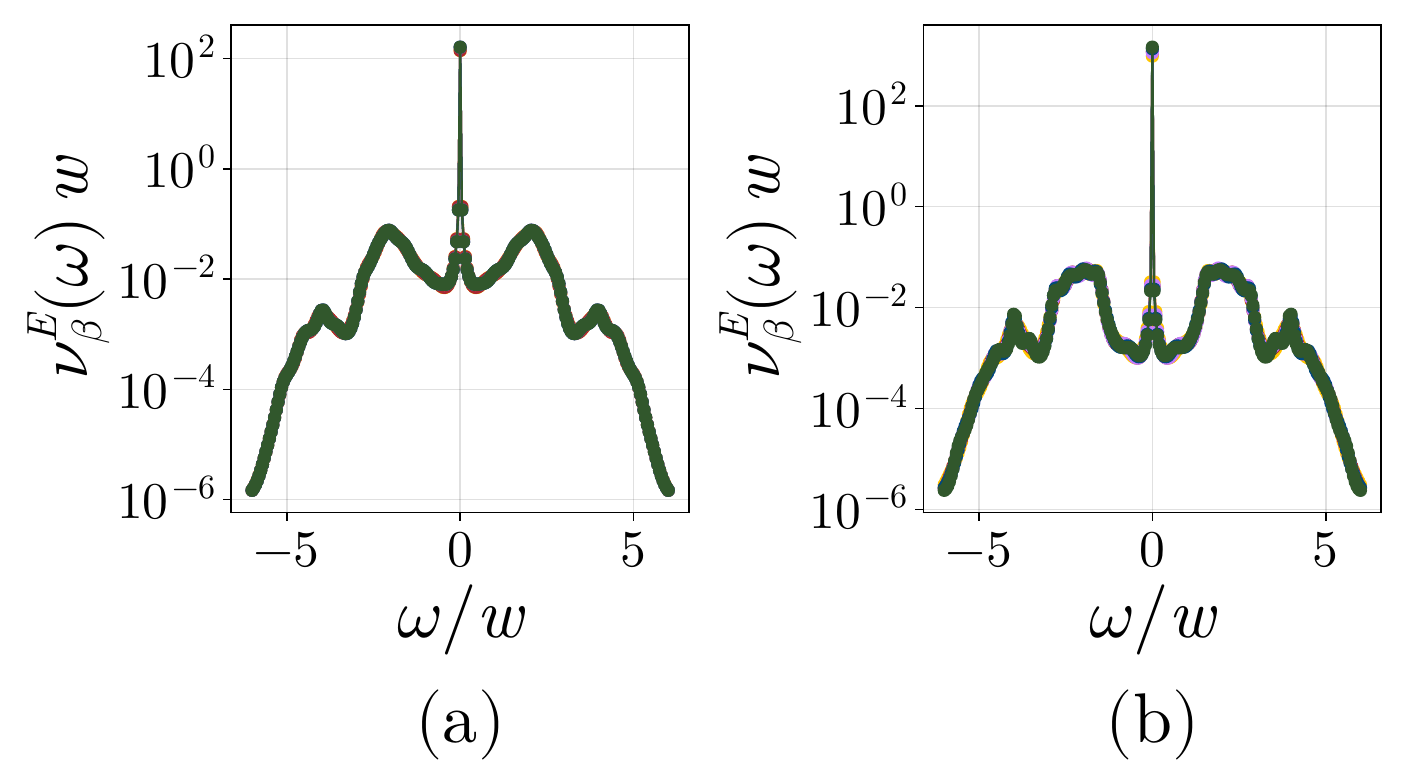}
    \caption{EDOS obtained from the Lanczos series $b_n$ from Fig.~\ref{fig:convergence_lanczos}, with the same identification of color and bond dimension.
    Panel (a) [(b)] is for $w\beta = 0.4$ ($w\beta = 1.25$).
    To obtain meaningful results, we discarded all Lanczos coefficients for $n$ larger than $n_{\rm unstable}$ as explained in the text.
    The collapse of all curves demonstrates the convergence of the EDOS with respect to the bond dimension.}
    \label{fig:convergence_edos}
\end{figure}
In Fig.~\ref{fig:convergence_edos} we plot the EDOS obtained from the Lanczos series in Fig.~\ref{fig:convergence_lanczos}, by removing all $b_n$ for $n>n_{\rm unstable}$.
We observe that, unlike the Lanczos series, the EDOS shows little dependence on the bond dimension.
In addition, the central peak converges rapidly in the bond dimension $\chi$, and only small changes occur to the side bands.

This also implies that the lifetime $\gamma(\beta)$ obtained by fitting the central Lorentzian converges rapidly in $\chi$, while only the short-time dynamics is modified by increasing the bond dimension.

\begin{figure}[htb!]
    \centering
    \includegraphics[width=0.8\linewidth]{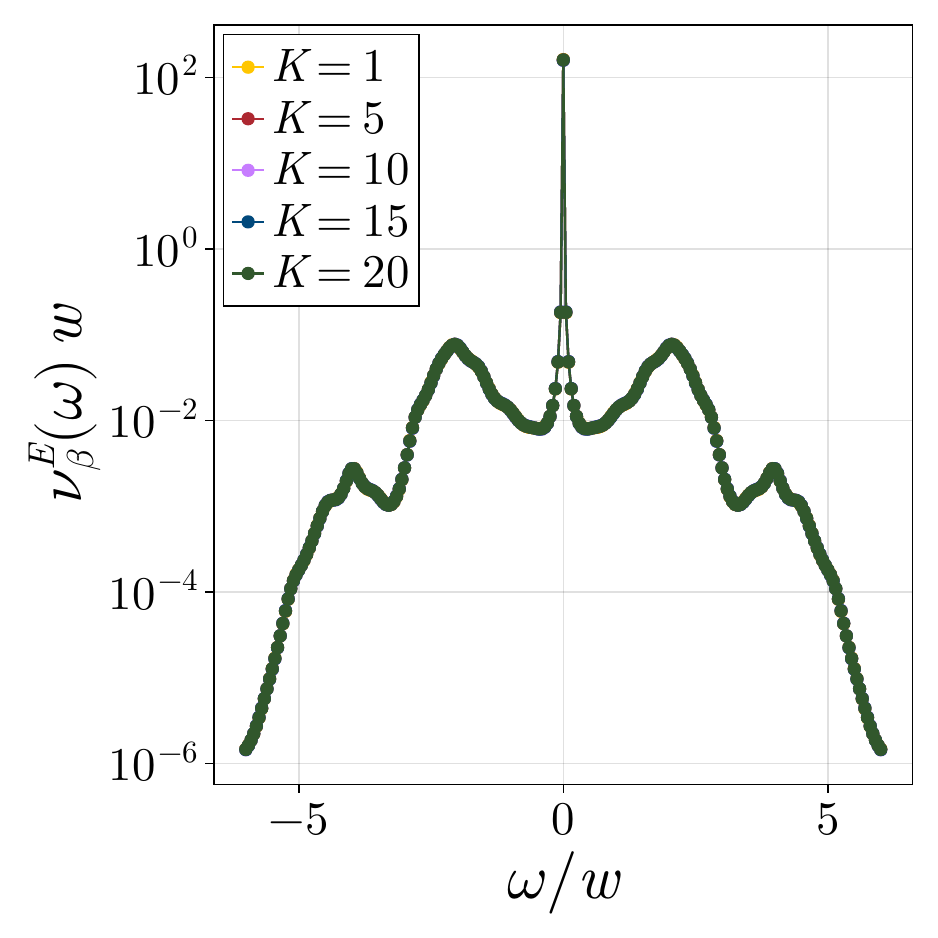}
    \caption{
        EDOS obtained by varying the hopping strength $w$ attaching the semi-infinite lead to the finite chain, with the latter described by the numerically obtained Lanczos coefficients $b_n$. The different hopping parameters $w$ are obtained by varying the averaging window $K$ according to equation~\eqref{eq:lead_attach_hopping}.
        The Lanczos coefficients are obtained for $\mu/w = 0/6$,\ $U/w = 0.3$, and $w\beta = 0.4$.
    }
    \label{fig:lead_attachments}
\end{figure}

We now discuss the second approximation, that involving the choice of the hopping parameter $w$ for the semi-infinite chain.
In all our results we have chosen $w = ib_N$, which is the last Lanczos coefficient before the instability sets in.
Alternately, one could have averaged over the last $K$ coefficients
\begin{equation}\label{eq:lead_attach_hopping}
    w = i \frac{1}{K} \sum_{k=0}^{K-1} b_{N-k} \, .
\end{equation}

In Fig.~\ref{fig:lead_attachments} we show the EDOS for different window sizes $K$.
Since all the curves lie on top of each other, we conclude that the explicit value of $w$ is not that important as long as $w$ faithfully represents the plateau value.

\begin{figure}[htb!]
    \centering
    \includegraphics[width=\linewidth]{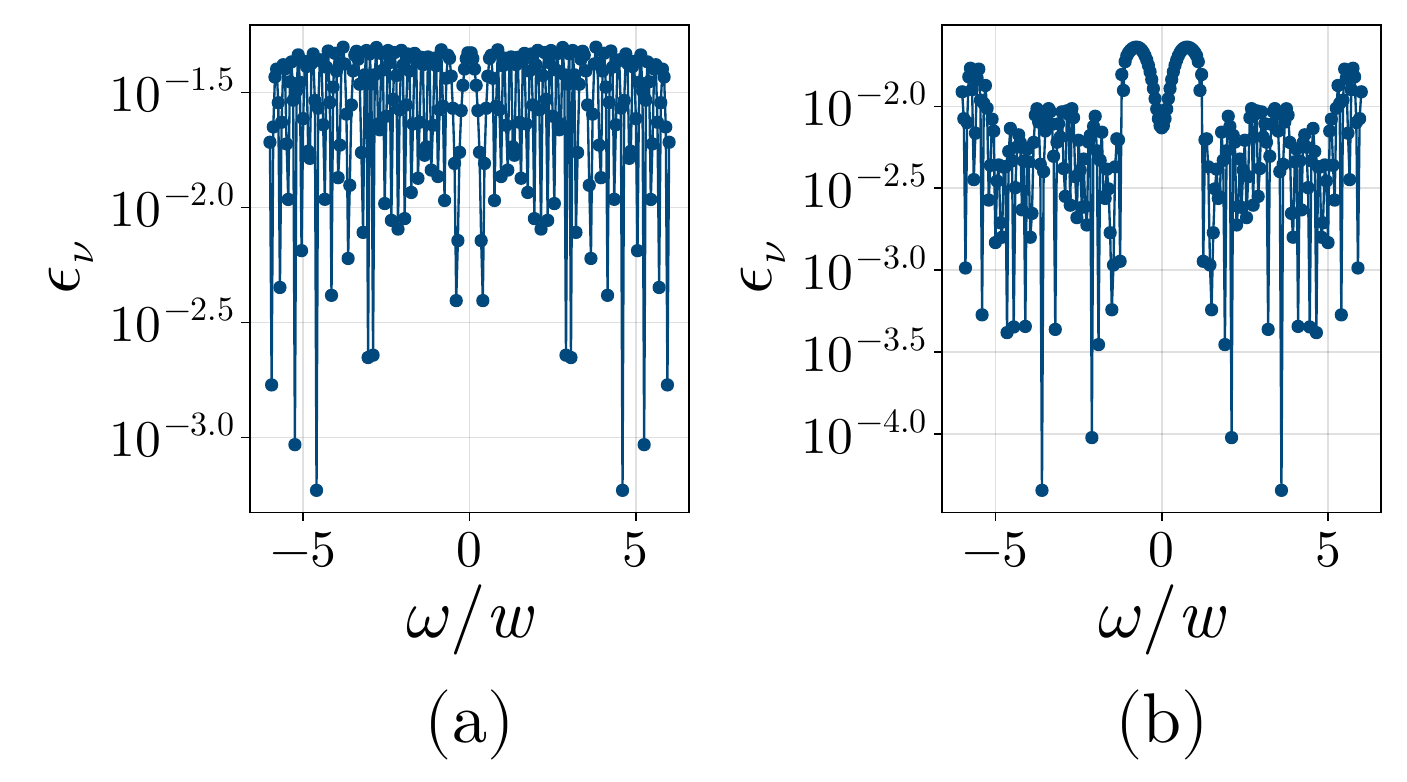}
    \caption{
       Energy resolved relative error $\epsilon_{\nu}$ of the EDOS, see Eq.~\eqref{eq:relative_error_dos}. Panel (a) [(b)] shows the error for $w\beta = 0$ ($w\beta = 2.45$). 
    }
    \label{fig:dos_relative_error}
\end{figure}

    Finally, we demonstrate the convergence of our results with respect to the system size $L$.
    For this, we calculate the EDOS for the two system sizes $L_1 = 22$ and $L_2 = 30$ from which we extract the relative error
    \begin{equation}\label{eq:relative_error_dos}
                \epsilon_{\nu}(\omega, \beta) \coloneqq 
                \left|\frac{\nu^E_\beta(\omega, L_1) - \nu^E_\beta(\omega,L_2)}{ \nu^E_\beta(\omega,L_2)}\right|\,.
    \end{equation}
    Fig.~\ref{fig:dos_relative_error} shows this relative error for $\mu/w = 1.2$, $U/w = 0.1$ and the two inverse temperatures $w\beta = 0.0$ and $w\beta = 2.35$.
    In both cases, the relative error is of order $10^{-2}$ for all energy scales.
    This is also reflected in the width parameter
    \begin{equation}\label{eq:relative_error_width}
        \begin{split}
            {}&\left|\frac{\gamma(0,L_1) - \gamma(0,L_2)}{\gamma(0,L_2)}\right| \approx 0.05 \\
            {}&\left|\frac{\gamma(2.35,L_1) - \gamma(2.35,L_2)}{\gamma(2.35,L_2)}\right| \approx 0.007 \,.
        \end{split}
    \end{equation}
    As this study is not aiming to achieve the most precise values and due to the growth of computational resources with system size, we present results for $L = 22$ as this system size is already sufficient for extracting reasonable numbers. A systematic study of the influence of the system size is left for future work.

\subsection{\label{sec:ap:wightman_vs_standard}Wightman -- Standard scalar product}

\begin{figure}[htb!]
    \centering
    \includegraphics[width=0.8\linewidth]{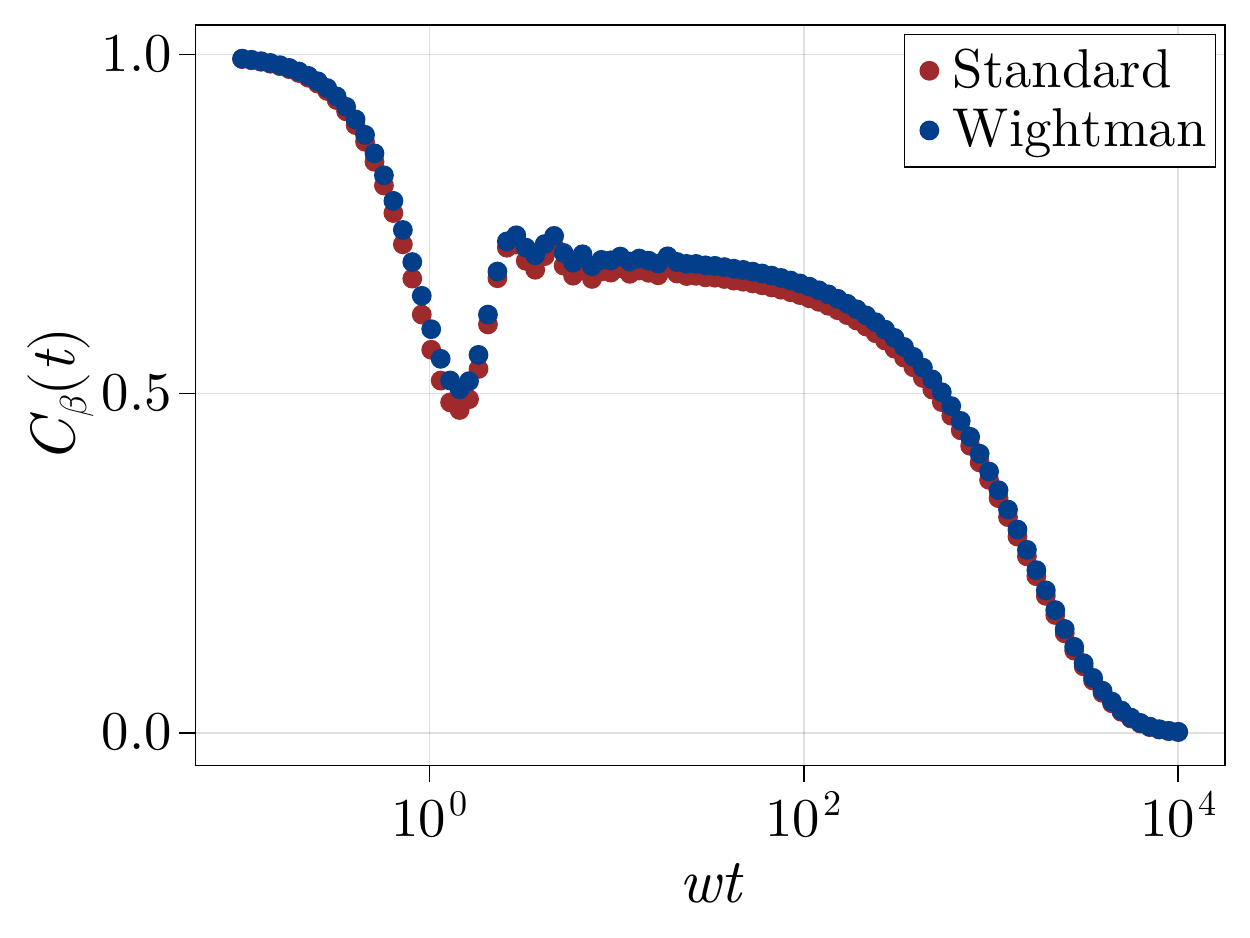}
    \caption{
    Comparison between the autocorrelation function obtained using the scalar product (red) and the Wightman scalar product (blue).}
    \label{fig:wightman_vs_standard}
\end{figure}

In this appendix we demonstrate that the Wightman finite temperature scalar product defined in equation~\eqref{eq:basic_notions:autocorr_finite_temp} in appendix \ref{sec:ap:ftsclprd} gives qualitatively the same result as the standard scalar product which we have exclusively used throughout this paper.
In Fig.~\ref{fig:wightman_vs_standard} we plot the autocorrelation function obtained for $\mu/w = 1.2$ and $U/w = 0.1$ for both choices of the scalar product.
Both results are obtained for a system size of $L=22$, and for the inverse temperature $w\beta = 0.4$.
We see that both curves have the same overall behavior, although the standard scalar product has a smaller plateau value at intermediate times.
The lifetime $\tau$ at which the autocorrelation functions decay to zero is qualitatively the same for both choices of the scalar product.

\subsection{\label{sec:ap:tdvp_vs_lanczos}TDVP vs Lanczos}

\begin{figure}[htb!]
    \centering
    \includegraphics[width=\linewidth]{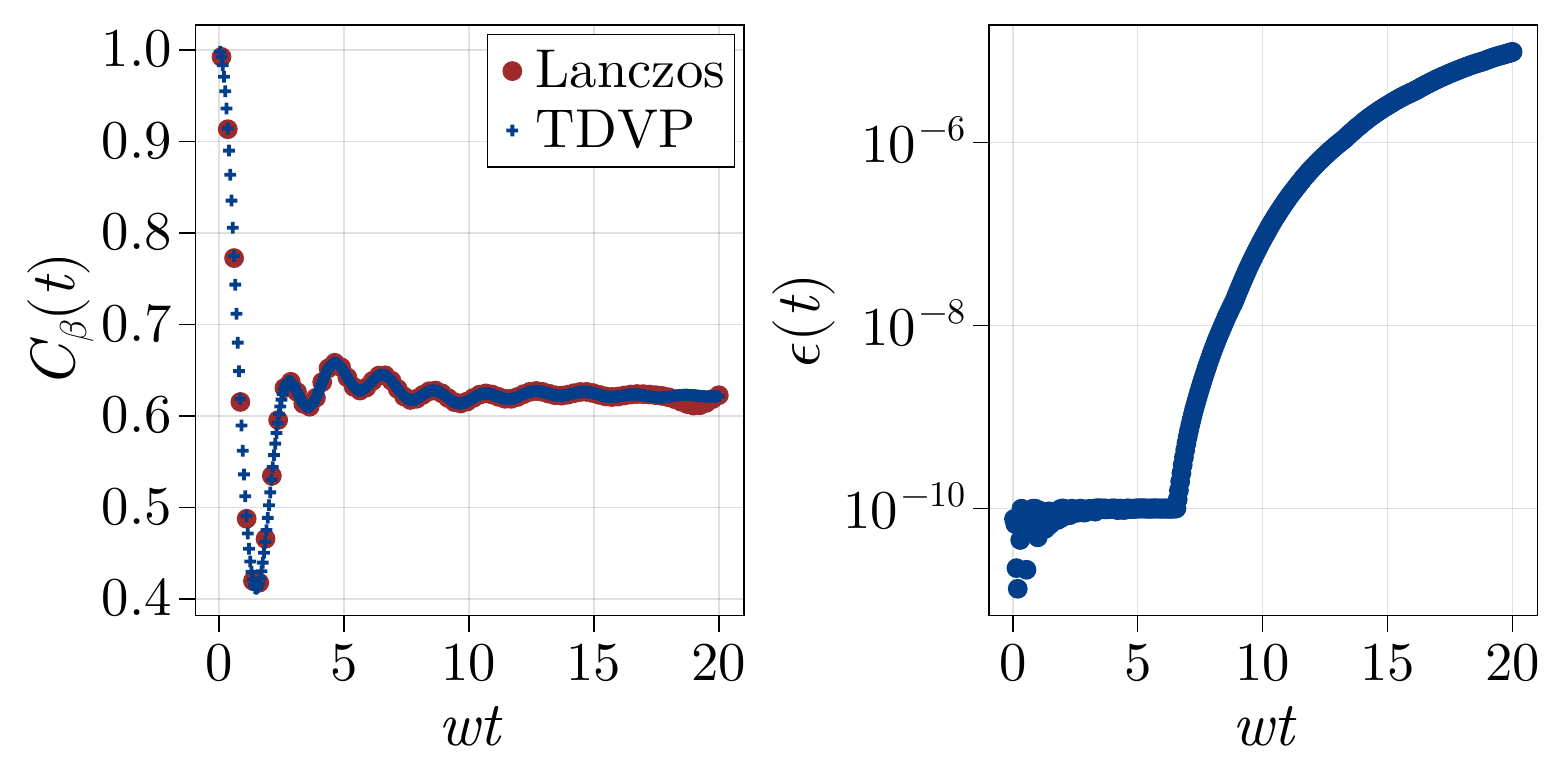}
    \caption{
    Comparison between the autocorrelation function obtained from the Lanczos series (red dots) and the TDVP algorithm (blue crosses) at infinite temperature in the topological region ($\mu/w = 1.2$, $U/w = 0.1$).
   Panel (a) shows the autocorrelation function. The left panel the truncated weight of the MPO of the TDVP simulation.}
    \label{fig:topo_tdvp}
\end{figure}

\begin{figure}[htb!]
    \centering
    \includegraphics[width=\linewidth]{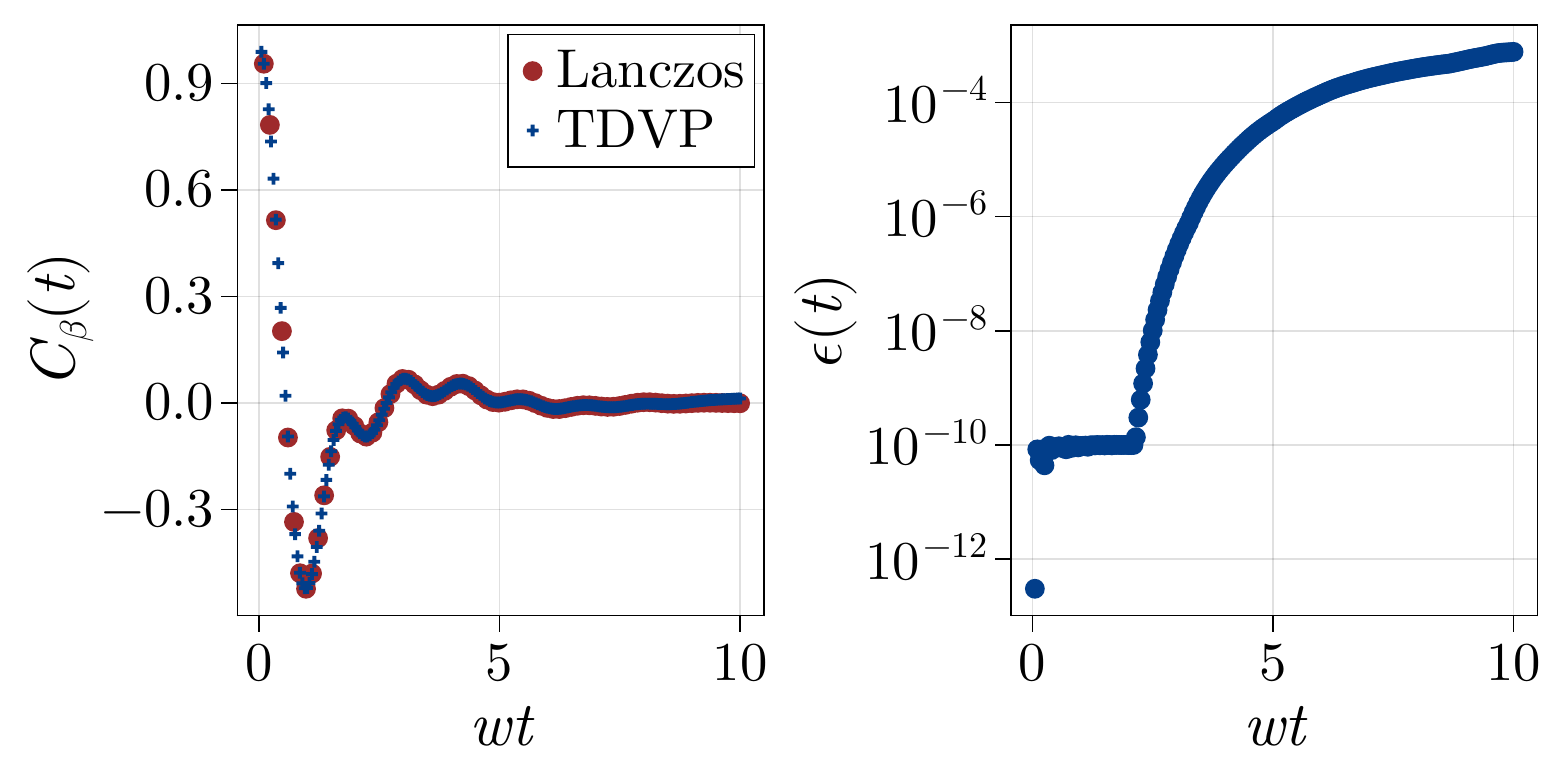}
    \caption{Comparison between the autocorrelation function obtained from the Lanczos series (red dots) and the TDVP algorithm (blue crosses) at infinite temperature in the Mott insulating region ($\mu/w = 0.2$, $U/w = 1.5$). Meaning of the panels are the same as in Fig.~\ref{fig:topo_tdvp}.}
    \label{fig:mott_tdvp}
\end{figure}

\begin{figure}[htb!]
    \centering
    \includegraphics[width=\linewidth]{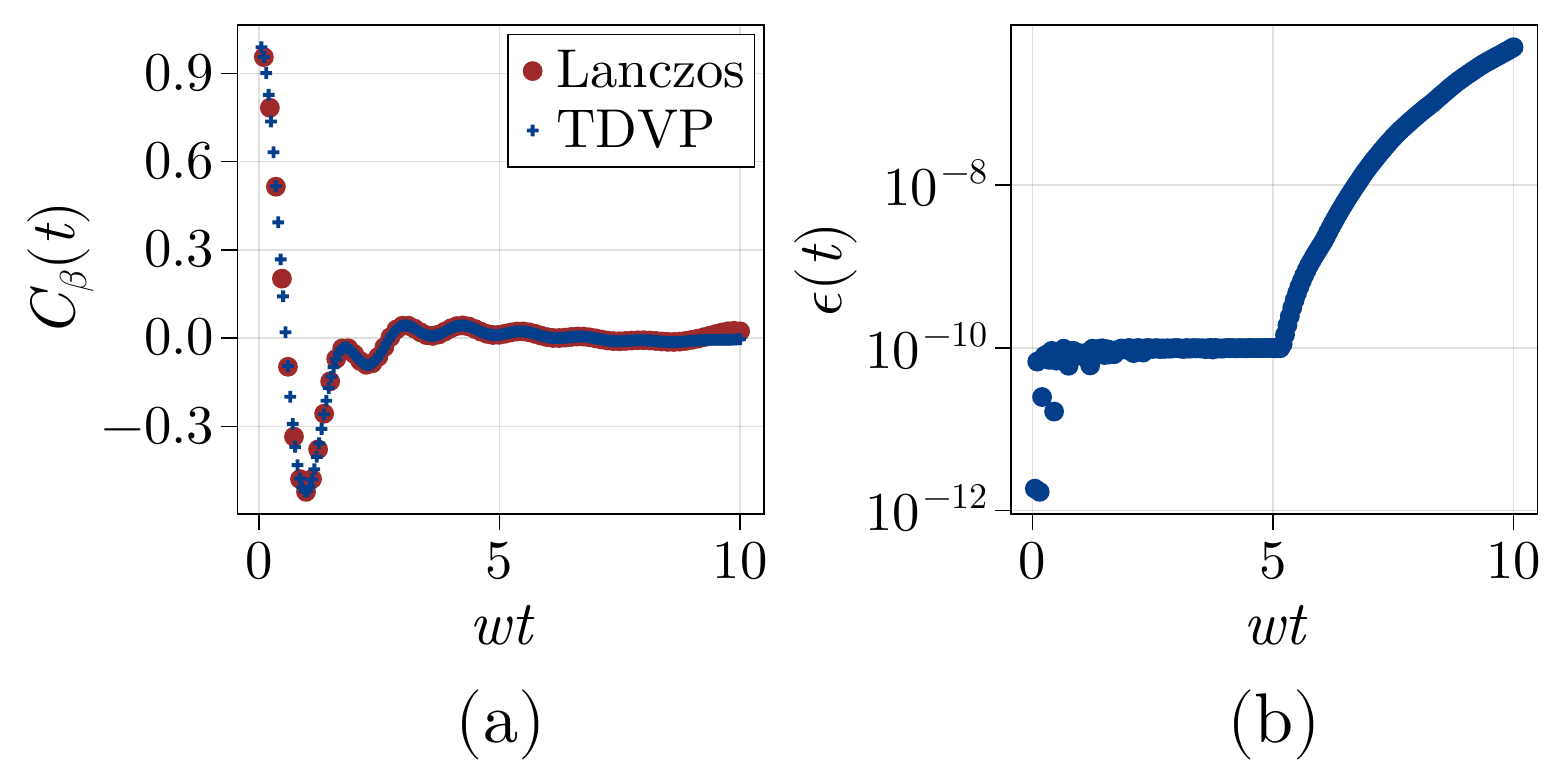}
    \caption{Comparison between the autocorrelation function obtained from the Lanczos series (red dots) and the TDVP algorithm (blue crosses) at infinite temperature in the trivial insulating region ($\mu/w = 3$, $U/w = 0.1$). Meaning of the panels are the same as in Fig.~\ref{fig:topo_tdvp}.}
    \label{fig:trivial_tdvp}
\end{figure}

In this appendix we compare the autocorrelation function obtained from the Lanczos series with direct integration of the Heisenberg time evolution using the time dependent variational principle (TDVP)~\cite{Haegeman2011, Haegeman2016}.
In the TDVP algorithm, we limit the bond dimension of the time evolved operator $\gamma_{1,a}(t)$ to $\chi = 512$.
To compare the TDVP with the Lanczos algorithm, we consider three different parameter points: 1) $\mu/w = 1.2$, $U/w = 0.1$, 2) $\mu/w = 0.1$, $U/w = 1.5$, and 3) $\mu/w = 1.5$, $U/w=0.1$.
For all three points we fixed $L = 22$ and considered infinite temperature ($\beta = 0$). For 1) we fixed the bond dimension of the Lanczos series to $\chi = 1500$, while for 2) and 3) we used $\chi = 1000$.
The first point is in the topological region of the model, see Fig.~\ref{fig:phasediagram_sketch}, while 2) and 3) are in the Mott insulating and trivial insulating phases, respectively.
The results are shown in Fig.~\ref{fig:topo_tdvp}--\ref{fig:trivial_tdvp}.

In general, we observe that the autocorrelation functions are consistent, even in the trivial and Mott insulating phase, where the correlation function decays rapidly.
However, at late times we observe deviations of the TDVP results from the Lanczos series.
This can be partially explained by the finite bond dimension of the MPO used in the TDVP algorithm, which leads to an error that increases with the simulated time $t$.
However, since the discarded weight, right panles of the figures~\ref{fig:topo_tdvp}--\ref{fig:trivial_tdvp}, is relatively small up to the considered times, we expect this not to be the only effect.

Another aspect may be the finite size of the system itself.
In the Lanczos algorithm, we approximated the unknown coefficients with a semi-infinite lead. This artificially extends the system to an infinite size, removing any kind of finite size effects from reflected excitations.
In the TDVP approach, we cannot consider an infinitely extended chain and thus the deviation might originate from these finite size effects.

We would also like to point out that the time required for the TDVP simulation is linear in the time steps and thus proportional to the final time.
For the Lanczos simulation, this is not the case, since we only need to extract a few hundred coefficients to obtain a fairly good result for the time evolution, even at late times.
This is especially important for the topological regime where we would like to access very large times $wt \sim  10^3$ in order to extract the lifetime of the ASZM.
For example, the TDVP simulation in Fig.~\ref{fig:topo_tdvp} needed $12$ hours on a 
\textrm{AMD EPYC 74F3} with $24$ cores.
In the last steps, the time consumption per TDVP iteration saturated around $700$ seconds.
A direct interpolation would give an estimate of $\sim 80$ days to complete a TDVP simulation with $\chi = 512$ up to times of $wt \sim  10^3$.
Moreover, it is certainly necessary to increase the bond dimension of the TDVP ansatz for these long times in order to obtain reasonable results.
For comparison, the Lanczos series with $\chi = 2000$ (used in Fig.~\ref{fig:topo_tdvp}--\ref{fig:trivial_tdvp}) took only three days to compute.

\subsection{\label{sec:ap:gap_extraction}Gap Extraction}

\begin{figure}
    \centering
    \includegraphics[width=\linewidth]{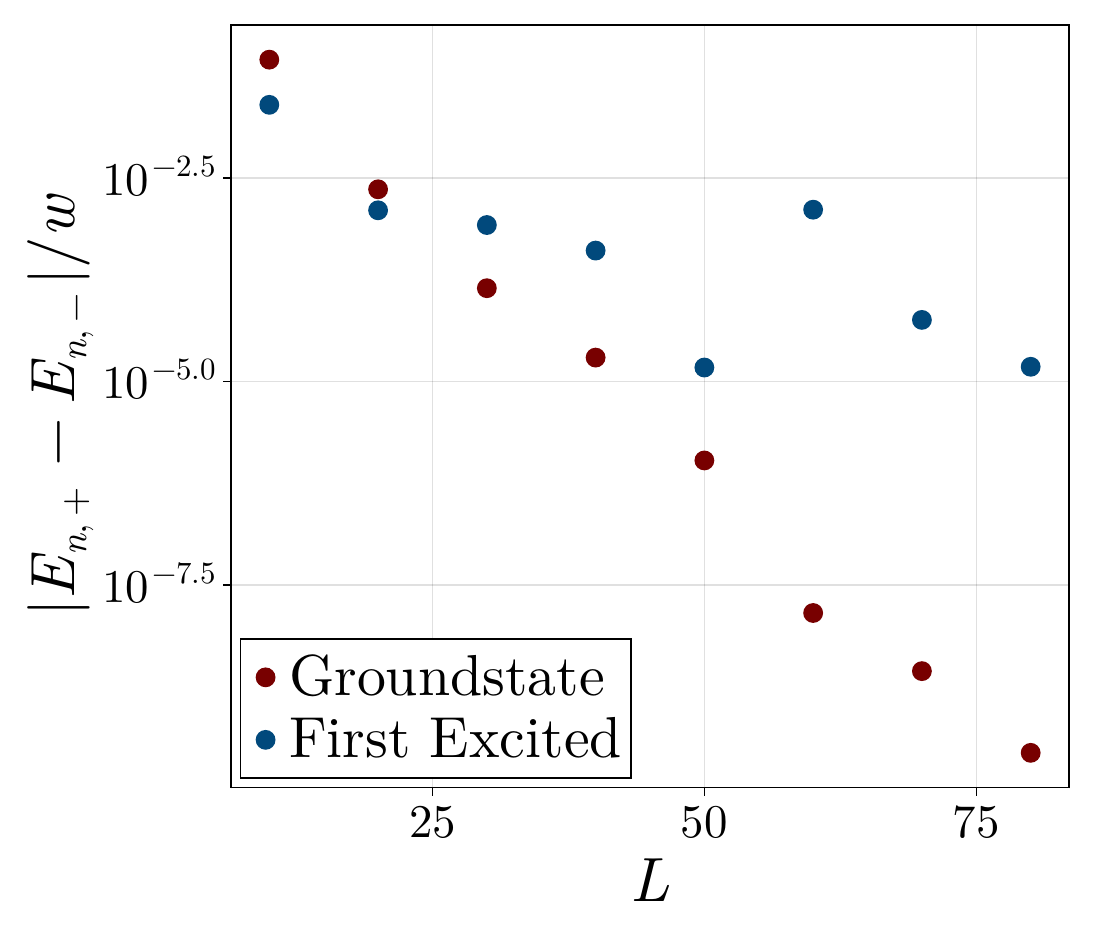}
    \caption{Energy gaps between different parity sectors for $\mu/w = 1.2, U/w = 0.6$, and for various system sizes $L$. The energy difference between the two ground states of opposite parity, vanishes exponentially with the system size. Similarly, the energy difference between the two excited states of opposite parity, vanishes exponentially for smaller system sizes. The noise for larger system sizes can be reduced by increasing the bond dimension.}
    \label{fig:energy_dmrg}
\end{figure}

We extracted the gaps of the many-body spectrum of the Kitaev-Hubbard model~\eqref{eq:kitaev_hubbard_chain} using the density matrix renormalization group (DMRG)~\cite{White1992, Schollwöck2011}.
With the DMRG we extracted the ground state of the Hamiltonian for the even and odd parity sectors, together with the first excited states within each parity sector.
For the extraction of the states we chose a matrix product state ansatz with a maximal bond dimension of $\chi = 200$. This allows for a good enough estimator of the gap in the thermodynamic limit.
However, we observe that for larger system sizes the ansatz is not sufficient to capture the first excited states with very high precision, see Fig.~\ref{fig:energy_dmrg}.

From the eigenstates $\ket{\psi_{n,p}}$, one can obtain the energy for every system size $L$ 
\[
    E_{n,p}(L) = \braket{\psi_{n,p}|H|\psi_{n,p}}\,.
\]
In Fig.~\ref{fig:energy_dmrg} we plot the gap $ \coloneqq |E_{n, +1} - E_{n, -1}|$ between opposite parity sectors of the ground states $n=0$, and the first excited states $n=1$, showing that this gap vanishes exponentially in the system size $L$.

\begin{figure}
    \centering
    \includegraphics[width=\linewidth]{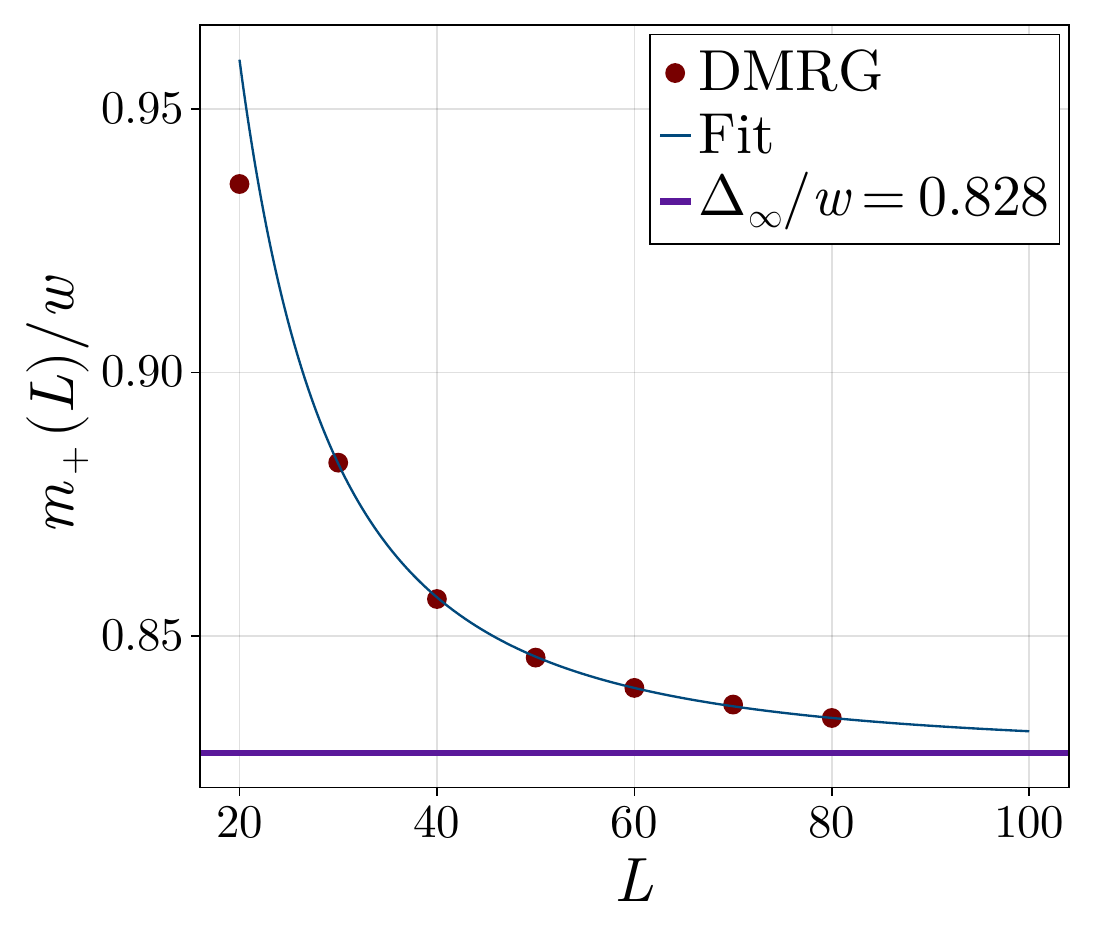}
    \caption{Energy gap between the ground state and the first excited state in the even parity sector (red dots) for $\mu/w = 1.2, U/w = 0.6$. The blue line is a fit to Eq.~\eqref{eq:finite_size_gap}
    and the horizontal purple line represents the thermodynamic limit of the gap $\Delta_\infty$ extracted from this fit.}
    \label{fig:energy_gap_dmrg}
\end{figure}

Next we consider the mass gap $m_{p}(L) = E_{1,p}(L) - E_{0,p}(L)$.
From general finite size scaling one expects
\begin{equation}
\label{eq:finite_size_gap}
m_{p}(L) = A/L^{\alpha} + \Delta_\infty,
\end{equation}
which can be fitted using linear regression in combination with an integral transformation~\cite{Jacquelin2009}.
In Fig.~\ref{fig:energy_gap_dmrg} we show that our extracted data shows the desired behavior. By fitting this algebraic decay, we can then extract the thermodynamic energy gap $\Delta_\infty$.

\section{\label{sec:ap:implementation_details}Details on the implementation of the truncation.}

The truncation of the application of $\liouv$ on the MPO $\mathcal{O}_n$ is implemented by truncating the reduced density matrices~\cite{McCulloch2007} of the full network $\liouv \mathcal{O}_n = H \mathcal{O}_n - \mathcal{O}_n H$.
This can easy be understood by rewriting the MPO as a MPS by combining the domain and codomain into a single Hilbert space. 
This resembles the isomorphism on the local Hilbert space
\[
\ket{\sigma^\prime}\bra{\sigma} \to \ket{\sigma^\prime, \sigma} \coloneqq \ket{\tau},
\]
which is anti-linear in the ket space, and linear in the bra space.
We denote the MPS obtained from $\mathcal{O}_n$ by this isomorphism as $\ket{\mathcal{O}_n}$.
Under this isomorphism, the commutator $\liouv$ becomes a standard MPO with bond dimension $K$.

Let $\ket{\Phi} = \liouv \ket{\mathcal{O}_n}$ denote the MPS one formally obtains by a full contraction of the network given by the right hand side.
The maximal bond dimension of $\ket{\Phi}$ is $K*\chi$, with $\chi$ being the bond dimension of $\ket{\mathcal{O}}$ and we would like to determine the most optimal truncation of $\ket{\Psi}$ back to a bond dimension of $\chi$. 
For this, we start by calculating the reduced density matrix of the last site:
\[
\rho_L \coloneqq \Tr_{L-1}\left[\ket{\Phi}\bra{\Phi}\right]\, ,
\]
where $\Tr_{L-1}$ is the trace over all sites $1$ to $L-1$.
In graphical notation

\vspace{2ex}

\begin{center}
\begin{tikzpicture}[baseline, scale = 0.85]
        \node[aten] (rho) at (0.0,0) {$\rho_4$};
        \draw (rho) -- ($(rho) - (0, 0.6)$) {};
        \draw (rho) -- ($(rho) + (0, 0.6)$) {};

        \node at (0.6,0) {$=$};
        \node[bten] (phi1) at (1.5, 0.5) {$\Phi_1$};
        \node[bten] (phi2) at (2.5, 0.5) {$\Phi_2$};
        \node[bten] (phi3) at (3.5, 0.5) {$\Phi_3$};
        \node[bten] (phi4) at (4.5, 0.5) {$\Phi_4$};
        
        \node[bten] (phi1bar) at (1.5, -0.5) {$\Phi_1$};
        \node[bten] (phi2bar) at (2.5, -0.5) {$\Phi_2$};
        \node[bten] (phi3bar) at (3.5, -0.5) {$\Phi_3$};
        \node[bten] (phi4bar) at (4.5, -0.5) {$\Phi_4$};

        \draw (phi1) -- (phi2);
        \draw (phi2) -- (phi3);
        \draw (phi3) -- (phi4);
        \draw (phi1bar) -- (phi2bar);
        \draw (phi2bar) -- (phi3bar);
        \draw (phi3bar) -- (phi4bar);

        \path [out=110,in=250] (phi1) edge (phi1bar);
        \path [out=110,in=250] (phi2) edge (phi2bar);
        \path [out=110,in=250] (phi3) edge (phi3bar);
        
        \draw (phi4) -- ($(phi4) + (0, 0.6)$) {};
        \draw (phi4bar) -- ($(phi4bar) - (0, 0.6)$) {};

        \node at (5.1, 0) {$=$};

        \node[cten, fit=(phi4)(phi4bar), inner sep = 0] (env) at (6, 0) {$E_{3}$};

        \node[bten] (phi4) at (7, 0.5) {$\Phi_4$};
        \node[bten] (phi4bar) at (7, -0.5) {$\Phi_4$};
        
        \draw (phi4) -- ($(phi4) + (0, 0.6)$) {};
        \draw (phi4bar) -- ($(phi4bar) - (0, 0.6)$) {};
        \draw  (env.east |- 0,0.5) -- (phi4);
        \draw  (env.east |- 0,-0.5) -- (phi4bar);

        \node at (7.6,0) {.};
\end{tikzpicture}
\end{center}

\vspace{2ex}

This also defines the environment $E_{k}$ containing all the contractions to the left.
This environment can be computed iteratively based on the previous environment $E_{k-1}$
\[
    E_{k} = \sum_{\tau} (\Phi_{k}^\tau)^\dagger E_{k-1}^{\vphantom \tau}\Phi_{k}^\tau,
\]
where $\Phi_k^\tau$ is the MPS tensor on the $k$-th site.
This density matrix is now diagonalized, where we only keep at most $\chi$ eigenstates
\[
\rho = \tilde{U}_L^\dagger D \tilde{U}_L^\vdag \, ,
\]
with $\tilde{U}_L^\vdag \tilde{U}_L^\dagger= \unit$.

The matrix $\tilde{U}_L^\dagger$ can be seen as the most optimal projection of the physical Hilbert space to at most $\chi$ degrees of freedom and is the last tensor in our new truncated MPS.
We can now construct the two-site reduced density matrix, with the last site transformed into the virtual Hilbert space

\vspace{2ex}

\begin{center}
\begin{tikzpicture}[baseline, scale = 0.85]
        \node[aten] (rho) at (0.0,0) {$\rho_{3,4}$};
        \draw (rho) -- ($(rho) - (0, 0.6)$) {};
        \draw (rho) -- ($(rho) + (0, 0.6)$) {};

        \node at (0.6,0) {$=$};

        \node[bten] (phi3) at (2.5, 0.5) {$\Phi_3$};
        \node[bten] (phi3bar) at (2.5, -0.5) {$\Phi_3$};
        \node[bten] (phi4) at (3.5, 0.5) {$\Phi_4$};
        \node[bten] (phi4bar) at (3.5, -0.5) {$\Phi_4$};

        \node[cten, fit=(phi3)(phi3bar), inner sep = 0] (env) at (1.5, 0) {$E_{2}$};

        \draw (phi3) -- ($(phi3) + (0, 0.6)$) {};
        \draw (phi3bar) -- ($(phi3bar) - (0, 0.6)$) {};
        \draw  (env.east |- 0,0.5) -- (phi3);
        \draw  (env.east |- 0,-0.5) -- (phi3bar);
        \draw (phi3) -- (phi4);
        \draw (phi3bar) -- (phi4bar);
        \node[dten] (Udag) at (3.5,1.5) {$U_4^\dag$};
        \node[dten] (U) at (3.5,-1.5) {$U_4^\vdag$};
        \draw (phi4) -- (Udag);
        \draw (phi4bar) -- (U);
        \draw (Udag) -- ($(Udag) + (0, 0.6)$) {};
        \draw (U) -- ($(U) - (0, 0.6)$) {};

        \node at (4.1,0) {$=$};

        \node[diagten] (D) at (5,0) {$D$};
        \node[dten] (U) at (5,1) {$U_{3}^\vdag$};
        \node[dten] (Udag) at (5,-1) {$U_{3}^\dag$};

        \draw (D) -- (U);
        \draw (D) -- (Udag);
        
        \draw  ($(U.north) + (0.1,0)$) -- ($(U.north) + (0.1, 0.3)$);
        \draw  ($(U.north) - (0.1,0)$) -- ($(U.north) + (-0.1, 0.3)$);
        \draw  ($(Udag.south) + (0.1,0)$) -- ($(Udag.south) + (0.1, -0.3)$);
        \draw  ($(Udag.south) - (0.1,0)$) -- ($(Udag.south) + (-0.1, -0.3)$);

        \node at (5.6,0) {,};
\end{tikzpicture}
\end{center}

\vspace{2ex}

\noindent which is again diagonalized, followed by truncation of the eigenspace to at most $\chi$ states.

We can now continue by successively constructing the new reduced density matrix for the sites $L-2$ to $L$ by projecting the sites $L-1$ and $L$ onto the virtual Hilbert space. This density matrix is again diagonalized and truncated.
We iterate until we reach the last point of the chain, which we simply keep as the last tensor in the new truncated MPS.

Note that the entire iteration does not require a full contraction of $\liouv \ket{\mathcal{O}_n}$, but only the environments, and this can be computed efficiently.
Also, the whole procedure does not require an explicit transformation of the MPO into an MPS, and one can work directly with the MPO.

\section{\label{sec:ap:mpo_xyz_model}MPO representation of the Strong Zero Mode in the XYZ Model}

\begin{figure}[ht!]
    \centering
        \begin{tikzpicture}[baseline={([yshift=-.5ex]current bounding box.center)},inner sep=1mm]
		\pgfmathsetmacro{\dz}{2}
		\pgfmathsetmacro{\dx}{1.8}
		
		\node[graphnode] (I)  at (0,0) {$0$};
		\node[graphnode] (F)  at (0,-\dz) {$3$};
        \node[graphnode] (1)  at (-\dx,-0.5*\dz) {$1$};
        \node[graphnode] (2)  at (\dx,-0.5*\dz) {$2$};

		\path[graphedge] (I) edge (F);
		
		\path[graphedge, every loop/.style={min distance=10mm,in=60,out=120,looseness=20}] (I) edge [loop above] node[above] {$J_zJ_y\unit$} (I);	
		\path[graphedge, every loop/.style={min distance=3mm,in=300,out=240,looseness=10}] (F) edge [loop below] node[below] {$\unit$} (F);

        \path[graphedge, every loop/.style={min distance=3mm,in=240,out=180,looseness=10}] (1) edge [loop below] node[below, anchor=north east] {$J_z\unit$} (1);
        
        \path[graphedge, every loop/.style={min distance=3mm,in=-60,out=00,looseness=10}] (2) edge [loop below] node[below, anchor=north west] {$J_y\unit$} (2);

       \path[graphedge] (I) edge [bend right=40] node[above=2pt, xshift = -2pt] {$(J_z^2 - 1)\sigma^z$} (1);
       \path[graphedge] (1) edge [bend right=40] node[above, xshift = -2pt] {$J_y\sigma^z$} (I);
       \path[graphedge] (I) edge [bend left=40] node[above=2pt, xshift=6pt, yshift=1pt] {$(J_y^2 - 1)\sigma^y$} (2);
       \path[graphedge] (2) edge [bend left=40] node[above, xshift = 8pt] {$J_z\sigma^y$} (I);

       \node at (-0.23*\dx,-0.7*\dz) {$\mathcal{N}\sigma^x$};
	\end{tikzpicture}
    \caption{Finite state machine representation of the SZM of the XYZ model.}
    \label{ap:fig:fsm_xyz_szm}
\end{figure}
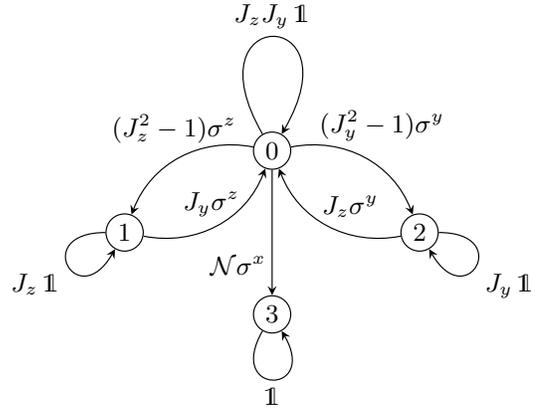

In Ref.~\cite{Fendley2016}, Fendley showed that the (integrable) XYZ model without a magnetic field (i.e., Eq.~\eqref{eq:xyz_ham} with $g =0$) hosts a SZM.
Here, we present a rewriting of the original rather complicated expression in terms of a MPO with bond dimension four and simple polynomial coefficients%
~\footnote{In a private communication, Paul Fendley mentioned a more general family of non-trivial MPOs of bond dimension-four commuting with the XYZ Hamiltonian, of which the SZM is a special case.}.
Without loss of generality, we set $J_x = 1$ and assume $J_y,J_z< 1$:
the SZM is then
\begin{equation}
    \Gamma = 
    \left(M^{0,\alpha_1}_{\kappa'_1,\kappa_1} 
M^{\alpha_1,\alpha_2}_{\kappa'_2,\kappa_2} 
    \dots
    M^{\alpha_{N-1}, 3}_{\kappa'_N,\kappa_N}\right)
    \ket{\lbrace\kappa'_j\rbrace}\bra{\lbrace\kappa_j\rbrace},
\end{equation}
with the operator-valued matrix $M^{\alpha, \alpha'}$ being
\begin{equation*}
    M = \begin{pmatrix}
	J_zJ_y \unit & (J_z^2 - 1)\sigma_z & (J_y^2 - 1)\sigma_y & \mathcal{N} \sigma_x\\
	0 & J_z\unit & 0 & J_y \sigma_z\\
	0 & 0 & J_y \unit & J_z \sigma_y\\
	0 & 0 & 0 & \unit
    \end{pmatrix}\,,
\end{equation*}
and $\mathcal{N}^2 = (1 - J_y^2)(1-J_z^2)$.
A graphical representation in terms of a finite-state machine is displayed in Fig.~\ref{ap:fig:fsm_xyz_szm}.
For $J_z=J_y=0$, it reduces to the familiar $\Gamma=\sigma^x_1$ for the Ising model.

\section{\label{sec:ap:exact_diagonalization}Construction of Strong Zero Mode from spectral data}

In this appendix, we review the construction of an (almost) strong zero mode operator $\szm$ from the full set of eigenstates and eigenenergies of the Hamiltonian $H$ \cite{Kells15}.
Because of the parity symmetry, all eigenstates $\ket{\psi_{n,p}}$ and eigenenergies $E_{n,p}$ are labeled by their associated parity $p = \pm 1$.
A general Majorana operator can now be written in this eigenbasis as
\[
\szm = \sum_{n,m} g_{n,m} \ket{\psi_{n,+1}}\bra{\psi_{m,-1}} + \hc\,,
\]
where $g\in U(\dim(\hilb)/2)$ is a unitary matrix and $\hilb$ is the underlying Hilbert space of the problem.
This operator anti-commutes with the parity operator and is hermitian $\szm^\dag = \szm$.
Further, one has
\[
\begin{split}
\szm^2 = \sum_{n,m} &(g\,g^\dag)_{m,n} \ket{\psi_{m,+1}}\bra{\psi_{n,+1}} \\
    {}+ &(g^\dag g)_{m,n} \ket{\psi_{m,-1}}\bra{\psi_{n,-1}} = \unit,
\end{split}
\]
where the last equality follows from the $g$ being a unitary matrix.
Plugging this ansatz into $\norm{[\szm, H]}^2$, one finds
\[
\norm{[\szm, H]}^2 = \sum_{n,m} |g_{n,m}|^2 (\epsilon_n - \epsilon_m)^2 \, .
\]
Assuming no degeneracies, this is minimized by $g_{mn} = g_n \delta_{n,m}$ with $g_n \in \U(1)$ being an arbitrary phase.
The fact that the operator is only specified up to some phases also reflects the gauge freedom to redefine states by arbitrary phases $\ket{\tilde{\psi}_{n, p}} = \exp(i\varphi_{n,p})\ket{\psi_{n,p}}$.
To obtain a gauge independent result, the phases $g_n$ must transform in a certain way when the gauge is changed:
\[
\tilde{g}_{n} = e^{-i(\varphi_{n,+1} - \varphi_{n,-1})} g_n \, .
\]
We now fix the gauge of each eigenstate by calculating the argument of the overlap with the operator $\gamma_{1,a} = c_1^\vdag + c_1^\dag =X_1$
\[
    \alpha_{n} \coloneqq \mathrm{Arg}(\braket{\psi_{n,+1} | X_1 | \psi_{n,-1}}),
\]
and redefining the states as:
\[
    \ket{\tilde{\psi}_{n,+1}} = e^{i\alpha_n}\ket{\psi_{n,+1}},\,\  \ket{\tilde{\psi}_{n,-1}} = \ket{\psi_{n,-1}}\,.
\]
By fixing the phase of the eigenvectors in this way, the transition operator $\ket{\tilde{\psi}_{n,+1}}\bra{\tilde{\psi}_{n, -1}}$ becomes gauge independent and the Majorana operator minimizing the commutator with the Hamiltonian reads
\begin{equation}
    \label{eq:ap:szm_ed}
    \szm = \sum_n g_n \ket{\tilde{\psi}_{n,+1}}\bra{\tilde{\psi}_{n, -1}} + \hc\,,\ g_n \in U(1)\, .
\end{equation}
For any choice of the phases $g_n$, Eq.~\eqref{eq:ap:szm_ed} defines a valid operator with minimal commutator.
This reflects the possibility to dress a solution with unitaries generated by a polynomial of the Hamiltonian $P(H)$:
\[
\tilde{\szm} \coloneqq e^{iP(H)}\szm e^{-iP(H)}\,.
\]
To find a unique solution, we fix the phases by requiring a maximal overlap with the operator $\gamma_{1,a}$.
This is achieved by setting $g_n = 1$ for all $n$. If the spectrum is perfect degenerate between the two parity sectors  $E_{n,p} = E_n$, the operator $\Gamma$ has a vanishing commutator with the Hamiltonian.
In this case, $\Gamma$ is a true zero mode.

\bibliography{refs.bib}

\end{document}